\documentclass[iop, twocolumn, numberedappendix, tighten]{emulateapj}

\usepackage{datetime}
\usepackage{graphicx}
\usepackage{dcolumn}
\usepackage{times}
\usepackage{amsmath}
\usepackage{enumerate}
\usepackage{natbib}
\usepackage{url}
\usepackage[utf8]{inputenc}
\usepackage{color}
\usepackage[colorlinks,urlcolor=blue,citecolor=cyan,linkcolor=red]{hyperref}

\newcommand{\beq}{\begin{equation}}
\newcommand{\eeq}{\end{equation}}
\DeclareMathOperator \dm {d}
\def\EAH{\textit{Einstein@Home}}

\newcommand{\bwswitch}{0}
\submitted{submitted to ApJ}

\shorttitle{\EAH{} Discovery of 24 PMPS Pulsars}
\shortauthors{Knispel et al.}

\begin{document}

\title{\EAH{} Discovery of 24 Pulsars in the Parkes Multi-beam Pulsar Survey}

\author{ B.~Knispel\altaffilmark{1,2,$\dagger$},
  R.~P.~Eatough\altaffilmark{3}, H.~Kim\altaffilmark{1,2,4},
  E.~F.~Keane\altaffilmark{3}, B.~Allen\altaffilmark{1,2,5},
  D.~Anderson\altaffilmark{6}, C.~Aulbert\altaffilmark{1,2},
  O.~Bock\altaffilmark{1,2}, F.~Crawford\altaffilmark{7},
  H.-B.~Eggenstein\altaffilmark{1,2}, H.~Fehrmann\altaffilmark{1,2},
  D.~Hammer\altaffilmark{5}, M.~Kramer\altaffilmark{3,8},
  A.~G.~Lyne\altaffilmark{8}, B.~Machenschalk\altaffilmark{1,2},
  R.~B.~Miller\altaffilmark{9}, M.~A.~Papa\altaffilmark{5,10},
  D.~Rastawicki\altaffilmark{7,11}, J.~Sarkissian\altaffilmark{12},
  X.~Siemens\altaffilmark{5}, and B.~W.~Stappers\altaffilmark{8}}

\altaffiltext{$\dagger$}{Email:
  \href{mailto:benjamin.knispel@aei.mpg.de}{benjamin.knispel@aei.mpg.de}}
  
\altaffiltext{1}{Albert-Einstein-Institut, Max-Planck-Institut f\"ur
  Gravitationsphysik, D-30167 Hannover, Germany}

\altaffiltext{2}{Institut f\"ur Gravitationsphysik, Leibniz
  Universit\"at Hannover, D-30167 Hannover, Germany}

\altaffiltext{3}{Max-Planck-Institut f\"ur Radioastronomie, D-53121
  Bonn, Germany}

\altaffiltext{4}{ Institut f\"ur Angewandte Physik, Universit\"at
  Bern, Sidlerstrasse 5, CH-3012 Bern, Switzerland}

\altaffiltext{5}{Physics Department, University of Wisconsin--Milwaukee,
  Milwaukee, WI 53211, USA}

\altaffiltext{6}{University of California at Berkeley, Berkeley, CA 94720 USA}

\altaffiltext{7}{Department of Physics and Astronomy, Franklin and
  Marshall College, P.O. Box 3003, Lancaster, PA 17604, USA}

\altaffiltext{8}{Jodrell Bank Centre for Astrophysics, School of
  Physics and Astronomy, University of Manchester, Manchester, M13
  9PL, UK}

\altaffiltext{9}{Department of Physics, West Virginia University, 111
  White Hall, Morgantown, WV 26506, USA}

\altaffiltext{10}{Albert-Einstein-Institut, Max-Planck-Institut f\"ur
  Gravitationsphysik, D-14476 Golm, Germany}

\altaffiltext{11}{Department of Physics, Stanford University, Stanford,
  CA 94305, USA}

\altaffiltext{12}{CSIRO Parkes Observatory, Parkes, NSW 2870,
  Australia}

\begin{abstract}
  We have conducted a new search for radio pulsars in compact binary
  systems in the Parkes multi-beam pulsar survey (PMPS) data,
  employing novel methods to remove the Doppler modulation from binary
  motion. This has yielded unparalleled sensitivity to pulsars in
  compact binaries. The required computation time of $\approx
  17\,000$\,CPU core years was provided by the distributed volunteer
  computing project \EAH{}, which has a sustained computing power of
  about 1~PFlop\,s$^{-1}.$ We discovered 24 new pulsars in our search, of
  which 18 were isolated pulsars, and six were members of binary
  systems. Despite the wide filterbank channels and relatively slow
  sampling time of the PMPS data, we found pulsars with very large
  ratios of dispersion measure (DM) to spin period.  Among those is
  PSR~J1748$-$3009, the millisecond pulsar with the highest known DM
  ($\approx$420\,pc\,cm$^{-3}$). We also discovered PSR~J1840$-$0643,
  which is in a binary system with an orbital period of 937 days, the
  fourth largest known. The new pulsar J1750$-$2536 likely belongs to
  the rare class of intermediate-mass binary pulsars. Three of the
  isolated pulsars show long-term nulling or intermittency in their
  emission, further increasing this growing family. Our discoveries
  demonstrate the value of distributed volunteer computing for
  data-driven astronomy and the importance of applying new analysis
  methods to extensively searched data.
\end{abstract}


\section{Introduction}
\label{sec:introduction}
\EAH{} is a distributed volunteer computing project: members of the
public donate otherwise unused computing cycles on their home and/or
office PCs to the project to enable blind searches for unknown neutron
stars. The detection of continuous gravitational waves from such
sources in data from ground-based interferometric detectors is the
main, long-term goal of \EAH{} \citep{2009PhRvD..79b2001A,
  2009PhRvD..80d2003A,2012arXiv1207.7176A}. Recently, successful
searches for neutron stars through their radio emission have also been
conducted with \EAH{} \citep{2010Sci...329.1305K,
  2011ApJ...732L...1K, 2013arXiv1303.0028A}.  Here we
present the first results from a novel search for radio pulsars in
compact binary systems in data from the Parkes multi-beam pulsar
survey (PMPS; \citealp{2001MNRAS.328...17M}).

The discovery and timing of compact binary pulsars is one of the key
science drivers for the construction of the Square Kilometre Array
\citep{2004NewAR..48..993K, 2004NewAR..48.1413C}. Such systems provide the most precise
`laboratories' for tests of general relativity and alternative
theories of gravity in the strong field regime
\citep{1989ApJ...345..434T, 2006Sci...314...97K, 2012CQGra..29r4007F,
  2012MNRAS.423.3328F, sw12}. The discovery of such systems also has
important implications for estimates of the Galactic binary merger
rate and for predictions of the expected detection rate of these
events in gravitational-wave searches, e.g.\ 
\citep{2003ApJ...584..985K}. Furthermore, the study of compact binary
pulsars offers unique opportunities to deepen our understanding of the
stellar evolution processes forming these systems
\citep{2004Sci...304..547S, 2008ApJS..174..223B} and their progenitor
stars. Finally, the discovery of pulsars hitherto missed in the PMPS
data helps to complete our picture of the Galactic pulsar population
and is useful for simulations for which the PMPS is used as a
`reference survey' \citep{2012arXiv1210.2746L}.

The detection of binary pulsars with standard Fourier methods is
hampered by Doppler smearing of the pulsed signal caused by binary
motion during the survey observation
\citep{1991ApJ...368..504J}. Previous searches for these systems in
the PMPS have utilized `acceleration searches' to correct for the
line-of-sight motion of the pulsars \citep{2004MNRAS.355..147F,
  ralphpaper, ralphthesis}. Although computationally efficient,
acceleration techniques are only effective when the observation time
is a small fraction of the orbital period and where the acceleration
is therefore roughly constant over the duration of the observation.
Thus, these techniques are not sensitive to the most compact systems
\citep{2002AJ....124.1788R}.

In this work, the PMPS data have been searched using a method to fully
demodulate the pulsar signals in compact binary systems based on a
large number of circular orbital templates. The method, which is only
possible with the computing resources provided by \EAH{}, offers
un-paralleled sensitivity to systems that would have previously been
missed by acceleration searches. Using tailored post-processing
methods to identify promising new pulsar candidates generated from the
search, we have discovered 24 new pulsars.

A full and detailed description of the \EAH{} radio pulsar search
pipeline can be found in \citet{2013arXiv1303.0028A}. The
goal of this paper is to provide a summary of the \EAH{} PMPS search
and to present our first discoveries. A sensitivity comparison to all
previous PMPS analyses, an estimate of the search sensitivity to
compact binaries, and implications for the Galactic population of
these objects will be discussed in a future paper.

The structure of the rest of this paper is as follows: in
Sec.~\ref{sec:earliersearches} we summarize the main system parameters
of the PMPS and previous analyses of the survey data.
Sec.~\ref{sec:pipeline} explains our new search pipeline in detail,
including post-processing methods. In Sec.~\ref{sec:discoveries} we
present the 24~newly discovered pulsars along with coherent timing
solutions in cases where they have been obtained. In
Sec.~\ref{sec:discussion} we briefly discuss implications of the
search, before concluding.  The reader can find technical details of
the post-processing in the Appendix.

\section{Parkes Multi-Beam Pulsar Survey and Previous
  Analyses}\label{sec:earliersearches}
The PMPS is the most successful pulsar survey ever performed with a
yield of over 800~pulsars. The success of the PMPS at discovering
pulsars is in part due to the multiple re-analyses of the data using
new and improved search methods. In searches for relativistic binary
pulsars, this has typically been enabled by an increase in the
available computing power, as in the case of this work. Here we
outline the survey setup, and previous and on-going analyses of the
PMPS data.

PMPS observations were done with the 64-meter Parkes radio telescope
and targeted the Galactic plane between $260^\circ$ and $50^\circ$
Galactic longitude $\ell$ and Galactic latitude $\left|b\right|
\leq5^\circ$ with a total of 3190~telescope pointings
\citep{2001MNRAS.328...17M}. Each pointing is comprised of
13~dual-polarization beams. These are arranged in two hexagonal rings,
each containing six beams around a central beam, creating a
`Star-of-David' pattern \citep{1996PASA...13..243S}. Each observation
covers a radio bandwidth of 288\,MHz, which is observed in a
filterbank of 96~channels each with 3\,MHz width, centered on
1374\,MHz. The sampling time is 250\,$\mu$s and the filterbank data
have a dynamic range of 1~bit per sample. Each beam has an integration
time of 2097.152\,s, with $2^{23}$ time samples. This yields a file
size of $\approx$100.7\,MB per beam and a total data volume of 4.1\,TB
for all filterbank and associated header files.

The majority of pulsars discovered in the PMPS ($\sim 600$) were found
in the original processing of the survey data
\citep{2001MNRAS.328...17M, 2002MNRAS.335..275M, 2003MNRAS.342.1299K,
  2004MNRAS.352.1439H}. After these analyses were completed, it was
noticed that in comparison to the number of typical isolated pulsars,
the number of binary and millisecond pulsars (MSPs) discovered was
disproportionally low \citep{2004MNRAS.355..147F}. For this reason a
full re-processing of the survey was performed with acceleration
searches, improved radio frequency interference (RFI) filters, and
better pulsar candidate selection tools \citep{2004MNRAS.355..147F}.
This analysis resulted in the discovery of a further 124~pulsars,
including 15~pulsars with spin periods $<30$\,ms. The double neutron
star system, PSR~J1756$-$2251, was also found and would not have been
discovered without acceleration searches \citep{2005ApJ...618L.119F}.
In addition, the results generated by the \citet{2004MNRAS.355..147F}
processing were re-analyzed using new tools for ranking of pulsar
candidates, allowing the discovery of another 29~pulsars
\citep{2009MNRAS.395..837K}.

Searches for binary radio pulsars can be characterized by the ratio of
phase-coherently analyzed observation time $T$ to orbital period
$P_\text{orb}$ of the pulsar. For orbital periods long compared to the
observation time, i.e.\ $T/P_\text{orb} \lesssim 0.1$, the signal can
be well described assuming a constant acceleration and `classical'
acceleration searches are a computationally efficient analysis method
\citep{2002AJ....124.1788R} with only small sensitivity losses. If
multiple orbits fit into a single observation, i.e., $T/P_\text{orb}
\gtrsim 5$, side-band searches provide a computational short-cut at
the cost of a slight loss in sensitivity \citep{2002A&A...384..532J,
  2003ApJ...589..911R}.

The acceleration search technique used by \citet{2004MNRAS.355..147F}
(stack-slide search) was imperfect in two respects. First, the
incoherent addition of spectra results in a reduction in sensitivity
(see, e.g., \citealp{1991ApJ...379..295W} and
\citealp{2000PhRvD..61h2001B}), and second, the orbital periods to
which the method was sensitive were limited $P_\text{orb} \gtrsim 10 T
\gtrsim 6$\,hr. These deficiencies were partially addressed by
\citet{ralphpaper}, where independent halves of the 35\,min
observation were analyzed with coherent acceleration searches,
resulting in a minimum detectable orbital period of $\sim3$\,hr. This
search, which also used improved RFI removal techniques and automated
pulsar candidate selection tools, resulted in the discovery of
16~pulsars, but no previously un-discovered relativistic binaries.

Recently, \citet{2012ApJ...759..127M} announced the discovery of five
MSPs in a re-analysis of the PMPS data. Because
no acceleration searches were used in this analysis, the reason these
pulsars were missed by previous searches is not yet fully understood.

The improved binary search analysis presented in this work was initiated
for two main reasons. Acceleration searches can only probe a limited
orbital parameter space. The data segmentation approach by
\citet{ralphpaper} is sensitive to a larger orbital parameter space but
loses sensitivity due to shorter coherent observation times. As we will
show below our search further expands the orbital parameter search while
using the full coherent observation time.

\section{The \EAH{} Analysis of the PMPS
  Data}\label{sec:pipeline}
The intermediate range $0.1 \lesssim T/P_\text{orb} \lesssim 5$ not
covered by acceleration or side-band searches is accessible at full
sensitivity by time-domain resampling with a large number of parameter
combinations for circular orbits. A widely used approach to this
problem in gravitational-wave data analysis is a matched filtering
process of convolving the data with multiple parameter combinations,
\citep{PhysRevD.53.6749, PhysRevD.60.022002, PhysRevD.76.082001,
  2009PhRvD..79b2001A, 2009PhRvD..80d2003A}.
  
The radio pulsar search with \EAH{} uses a time-domain re-sampling
scheme to search for radio pulsars in compact binary orbits
\citep{benthesis}. It features a newly developed, fully coherent
stage, which removes the frequency modulation of the pulsar signal
from binary motion in circular orbits.

After a summary of the \EAH{} project and the Berkeley Open Infrastructure
for Network Computing (BOINC) in Sec.~\ref{sec:boinc}, we describe the
data preparation methods in Sec.~\ref{subsec:preproc}.
In Secs.~\ref{subsec:sigmodel} to \ref{subsec:compaccsearch} we prepare a
theoretical background, and explain details of the search for
pulsars in binary orbits in Sec.~\ref{subsec:analysis}. Our post-processing
methods are described in Sec.~\ref{sec:postproc}.

\subsection{\EAH{} and The BOINC Framework}\label{sec:boinc}
The computing power for \EAH{} is provided by members of the general
public, donating idle compute cycles on their home and/or office
PCs. These highly fluctuating computing resources are managed using
BOINC software framework \citep{Anderson:2006:DRS:1188455.1188586} and a
set of central servers administered by a handful of project developers
and scientists.

BOINC is a framework to set up and manage a distributed volunteer
computing project that requires minimal attention from the volunteers
donating computing cycles. Members of the general public
(\textit{volunteers}) can register and attach their computers
(\textit{hosts}) to a BOINC project. For the volunteers, this only
requires downloading\footnote{\url{http://boinc.berkeley.edu/}} and
installing a small piece of software (\textit{BOINCManager}) and
selecting the distributed computing project of their
choice. Currently, there are several dozens of BOINC-based distributed
computing projects, covering a wide range of scientific
disciplines\footnote{\url{http://boinc.berkeley.edu/wiki/Project\_list}}.

The BOINCManager collects information about the host and then
coordinates the download of data files, analysis software, and
processing instructions. Only computational tasks (\textit{work
  units}) suitable for the given host are sent, and each work unit has
a certain deadline (2 weeks for \EAH{}) after which a result has to
be reported back to the project servers. Each work unit is sent to two
independent hosts, the results are then compared based on pre-defined
numerical metrics to ensure their scientific validity. If necessary,
additional copies of the work unit are sent to other independent hosts
until two agreeing results are found. A central database stores
information relevant to the work units, volunteers, hosts and internet
discussion forums. Details about the internal BOINC processes
controlling the work distribution as well as a more detailed
discussion of the type of problems that can be solved by distributed
volunteer computing projects are available in \citet{2013arXiv1303.0028A}.

\EAH{} is one of the largest distributed computing projects. Since
2005, more than 330\,000~volunteers have contributed to the
project. On average, about 48\,000~different volunteers donate
computing time each week on roughly 150\,000~different hosts.
Currently, the sustained computing power is of order 1 PFlop\,s$^{-1}$. The
radio pulsar search uses a varying fraction (between 20\% and 50\%) of
the central processing unit (CPU) time available to the project.  For
the radio pulsar search, additional computing time is available on
Nvidia and ATI/AMD graphics processing units (GPUs). Currently,
approximately 10\,000~hosts with Nvidia GPUs and 3600~hosts with
ATI/AMD contribute to the project each week. For the PMPS analysis
only executables for CPUs and Nvidia GPUs were available.

On average, 200 complete PMPS beams per day were analyzed by \EAH{}
between 2010~December and 2011~July, with three-day averages varying
between 170 and 230\,beams\,day$^{-1}$. The total computing time donated by
the volunteers to analyze the data set described here is of order
17\,000\,CPU core years.

\subsection{Pre-processing and De-dispersion}\label{subsec:preproc}
The PMPS data are publicly available and were copied from computer
systems at the Jodrell Bank Observatory to the Albert Einstein
Institute (AEI) in Hannover. For permanent storage at the AEI, data
were copied to a Hierarchical Storage Management System, backed by a
tape library. The filterbank data were also kept on spinning media to
provide low-latency availability for pre- and post-processing
purposes. In the following we describe the pre-processing and
de-dispersion applied to each of the 41\,364~observed beams in the
\EAH{} analysis.

The data were pre-processed on dedicated computers at the AEI
Hannover. The survey data are provided in `filterbank format', i.e.,
radio frequency power spectra resolved in `filterbank channels',
sampled at regular time intervals. In the first step, we convert
the survey data in filterbank format from 1 bit dynamic range per
sample into a file format with 8 bits per sample using tools from
the SIGPROC software toolbox\footnote{version 4.3 from
  \url{http://sigproc.sourceforge.net/}}.  This step is necessary for
further processing by other software from the
\textsc{presto}\footnote{git commit 95f0f4be23\dots from
  \url{https://github.com/scottransom/presto}} software toolbox
\citep{2002AJ....124.1788R} and does not add dynamic range. The
product of this first step is one filterbank file for each observed
beam.

To mitigate the effect of RFI at later stages in the analysis, we used
the \textsc{presto} software tool \textsc{rfifind} to obtain an
individual RFI mask for each beam. We analyzed the 8-bit filterbank
data in blocks of 2.048\,s length in each filterbank channel and
flagged persistent narrow-band RFI and transient broadband RFI.  In
each beam, up to a few percent of the data were identified as RFI in
this step. The next steps of the pre-processing replaced these blocks
by constant values.

The data were down-sampled by a factor of two in the time domain,
reducing the number of samples per time series to $2^{22}$ to save
computational time. This was achieved by co-adding
neighboring time series bins, increasing the sampling time to
$500$\,$\mu$s.

Free electrons in the interstellar medium delay the arrival time of
the radio waves with a dependence on their frequency.  If this effect
is not corrected for, it smears radio pulses observed over a wide
bandwidth, and severely reduces their detectability. The default
method to mitigate this effect is to incoherently de-disperse the
filterbank data, which is done by introducing frequency dependent
delays to the different filterbank channels \citep{pulsarhandbook}.
The time delay relative to a reference frequency depends on the a
priori unknown integrated electron column density along the line of
sight, the dispersion measure (DM).

For the de-dispersion with the \textsc{presto} tools, we used a set of
440~trial DMs up to 1876\,pc\,cm$^{-3}$, which exceeds the expected
range of DM values in our Galaxy \citep{2002astro.ph..7156C}. The set
of trial values is chosen using the \textsc{ddplan.py} tool from
\textsc{presto} and is shown in Tab.~\ref{tab:dmsteps}.

\begin{table}
  \caption{\rm Set of DM trial values used in the \EAH{} search of the PMPS data.}
  \setlength{\extrarowheight}{2pt}
  \label{tab:dmsteps}
  \begin{tabular*}{\columnwidth}{@{\extracolsep{\fill}}ccc}
    \hline
    \multicolumn{1}{c}{DM range} & \multicolumn{1}{c}{$\Delta \text{DM}$} & number of trial values\\
    \multicolumn{1}{c}{(pc\,cm$^{-3}$)} & \multicolumn{1}{c}{(pc\,cm$^{-3}$)} & \\
    \hline
    $0$ to $192$ & $1$ & $192$\\
	$192$ to $336$ & $2$ & $72$\\
	$336$ to $776$ & $5$ & $88$\\
	$776$ to $1436$ & $10$ & $66$\\
	$1436$ to $1876$ & $20$ & $22$\\
    \hline
  \end{tabular*}
\end{table}

We used \textsc{presto} tools to de-disperse, barycenter, and downsample the
filterbank data. The result of the de-dispersion step was a time series
encoded with 32 bits per sample, and an additional header file for each DM
trial value. A total of 440~de-dispersed time series and corresponding
header files was generated for each observed beam. Each of these
down-sampled time series was 16.8\,MB in size.

In the next step, we combined each de-dispersed time series and its
associated \textsc{presto} header file into a single file. The
external header files are included as file headers containing relevant
information about the time series data.

To save internet bandwidth for the data transfer to the \EAH{} hosts,
we encoded time series data with a dynamic range of 8~bits per sample.
The 8-bit dynamic range is sufficient to encode the whole possible
range of de-dispersed time series samples. Given the original 1-bit
sampling of the filterbank data and the number of filterbank channels
(96), the maximum possible value of any de-dispersed time series
sample is $96 \approx 2^{6.6}<2^8$.

Each `compressed' time series had a total size of 4.2\,MB. A bundle of
four time series from a given beam formed the input data for a single
\EAH{} work unit. A single modern CPU core with an approximate
computing power of $\sim$10\,GFlops\,s$^{-1}$ can analyze each task containing
16.8\,MB of data in $\approx$12\,hr.

The ratio of data I/O to computing time is therefore of order 1\,MB\,hr$^{-1}$,
which means that a 30\,MB\,s$^{-1}$ internet connection at a single download
server can keep of order $10^5$ hosts continuously busy, assuming
bandwidth is the limiting factor. Note that our search also runs on GPUs
which can finish the same task about
10-20~times faster than a CPU, see \citet{2013arXiv1303.0028A}
for details. For GPUs the I/O to computing time ratio increases to up
to 20 MB\,hr$^{-1}$.

A set of 400~pre-processed beams was kept ready for distribution at
any time.  This precautionary measure provided an ample time buffer
for the case of technical difficulties (e.g., with the pre-processing
machines) on the server side of the project.

\subsection{Signal Model and Detection
  Statistic}\label{subsec:sigmodel}
In searching for possible pulsar signals hidden in instrumental noise,
a signal model is required.  Detection statistics are then designed to
effectively identify that signal in detector noise.  A full
description of the signal model and detection statistics used in this
paper are given in \citet{2013arXiv1303.0028A}. Here, we
summarize the main points.

Our signal model describes the rotation phase $\Phi$ of the pulsar as
seen in the radio telescope at time $t$, assuming that the pulsar is
in a circular orbit: \beq \Phi\left(t \right) = 2\pi f\left(t +
  \frac{a\sin\left(i\right)}{c} \sin\left(\Omega_\text{orb} t +
    \psi\right)\right) + \Phi_0.\label{eq:fullphase} \eeq Here $f$ is
the intrinsic pulsar spin frequency, and $a\sin\left(i\right)$ is the
projected orbital radius with inclination angle $i$. The orbital
angular velocity $\Omega_\text{orb}$ is determined by the orbital
period $P_\text{orb}$ through
$\Omega_\text{orb}=2\pi/P_\text{orb}$. The angle $\psi$ denotes the
initial orbital phase and $\Phi_0$ is the initial signal phase.

To search for sinusoidal signals proportional to $\cos \Phi(t)$, a
large set of \textit{matched filters} is applied to the instrumental
output.  Each matched filter is optimized for a particular waveform,
and can be thought of as the ``best possible search'' for a signal at
a particular point in the parameter space.  The four parameters $f,
a\sin\left(i\right), \Omega_\text{orb}$, and $\psi$ are coordinates in this
parameter space of possible signals. The detection statistics can be
shown to be independent of the initial phase $\Phi_0$, cf.\ Sec.~4.3
of \citet{2013arXiv1303.0028A}.

The matched filter at a particular point in parameter space will also
respond to signals located ``nearby'', whose phase model is similar.
Thus, in constructing a computationally-efficient search, careful
consideration must be given to the set of filters is chosen. The set
of points in parameter space that is searched is called the
\textit{template bank}; Section~\ref{subsec:tbconstruct} explains how
it was selected.

At a particular point in signal parameter space, detection statistics are constructed
from the average power $\mathcal P_n$ in the $n$th harmonic of
the pulsar rotation phase $\Phi$. In signal processing, the $\mathcal
P_n$ are called ``matched filter squared signal-to-noise ratios''. This
assumes Gaussian white noise with unit variance in the data stream.
To search data for signals, the $P_n$ are computed for first 16 harmonics
of $\Phi$ (so $n=1,\dots,16$) and then combined to form detection
statistics.

The optimal way to combine the $\mathcal P_n$ into detection
statistics depends upon the pulsar's pulse profile.  For example if
the pulsar had a purely sinusoidal profile (at the rotation frequency)
then the $\mathcal P_1$ would be the optimal detection statistic.
However since we are searching survey data for \textit{new} pulsars,
we do not know the pulse profile ab initio. Thus, we search five
different detection statistics (often called \textit{harmonic sums})
$S_0, \dots, S_4$, constructed from summing different numbers of
harmonics together: \beq S_L \equiv \sum_{n=1}^{2^L} \mathcal P_n
.\label{eq:statdef} \eeq The $L$th harmonic sum is an optimal
detection statistic (in the Neyman-Pearson sense, maximizing the
detection rate for a given false-alarm rate) for a pulsar pulse
profile that is a Dirac delta-function spike, truncated at the
$2^L$th harmonic.

It is important to characterize the statistical properties of the
$S_L$.  In the absence of a pulsar, instrument noise can mimic a
signal, resulting in large values of the detection statistics and a
possible \textit{false alarm}.  If the instrument noise has Gaussian
statistics, then in the absence of a pulsar it is easy to see that
$S_L$ behaves like a classical $\chi^2$ random variable with
$2^{L+1}$ degrees of freedom. The \textit{false alarm probability}
$p_\text{FA}(S_L^*)$ is the probability that $S_L$ exceeds a threshold
value $S_L^*$ in the absence of a signal; for Gaussian noise this is
given by an incomplete upper $\Gamma$-function.

We use the \textit{significance} $\mathcal S$ to indicate the
statistical significance of a signal candidate.  Pulsar signals which
are strong enough to be observable produce unusually large values of
$\mathcal P_n$ and thus unusually large values of $S_L$.  These in
turn have low false-alarm probability. Hence we define the
significance of a candidate by \beq \mathcal S\left(S_L\right) \equiv
-\log_{10}\left(p_\text{FA}\left(S_L\right)\right).
\label{eq:sig}
\eeq For example, a candidate with significance $\mathcal S = 20$ has
probability $10^{-20}$ of occurring in random Gaussian noise.

\subsection{Template-bank Construction}
\label{subsec:tbconstruct}
The \textit{template grid} or \textit{template bank} is the set of
points in the parameter space at which the detection statistic is
evaluated.  In an ideal world (unlimited computing power) the template
bank would contain a very large number of signal templates: for any
signal in the parameter space, there would be a template located
nearby. No signal-to-noise ratio would be lost due to mismatch between
the signal and template parameters.

Real template banks are designed to maximize detection probability at
fixed computing cost. Substantial research work in the
gravitational-wave detection community has shown how to construct
(near-)optimal template banks (\citealp{PhysRevD.53.6749,
  PhysRevD.60.022002, 2009PhRvD..80j4014H, 2009PhRvD..79j4017M};
Fehrmann and Pletsch, in preparation).  A real template bank is
characterized by the its ``worst case'' mismatch.

The mismatch $m$ is defined (for a signal at one point in parameter
space, and a filter template at a different point) as the fractional
loss of detection statistic compared to a template placed at the
signal location. In this paper \textit{the same} template bank is used
for the detection statistics $S_0, \dots, S_4$, however the mismatch
we discuss is only that of the ``fundamental mode'' $S_0 = \mathcal
P_1$ detection statistic. The region of parameter space around a
particular template, whose mismatch is less than some nominal value is
called the template's \textit{coverage region}.  For small enough
nominal values ($m \lesssim 0.01$), the coverage region is an ellipsoid.
However for the worst-case mismatch used in this search (typically
$m=0.2$ or $m=0.3$) the region is ``banana-shaped'' as discussed in
Sec.~\ref{subsec:searchspace}.

Our template bank and its construction method are similar to those
described in \citet{2013arXiv1303.0028A} and
\citet{benthesis}. The four-dimensional template bank is obtained by a
Cartesian product of a three-dimensional \textit{orbital template
  bank} with a uniform frequency grid with spacing $\Delta f = 1/3T$.
The extra factor of $3$ is due to the mean padding used in our analysis,
see Sec.~\ref{subsec:analysis} for details.
The orbital template bank is not uniform. Since the region of constant
mismatch varies as one moves over the parameter space, and is not
ellipsoidal, the orbital template bank can not be a regular lattice,
constructed using standard methods such as
\citep{PhysRevD.60.022002}. We use a combination of the random
\citep{2009PhRvD..79j4017M} and stochastic \citep{2009PhRvD..80j4014H}
template bank constructions.  The orbital template bank was
constructed using about 200k CPU hours on the Atlas cluster
\citep{AtlasRef}.

\begin{figure}
  \begin{center}
    \ifcase \bwswitch
    \includegraphics[width=\columnwidth]{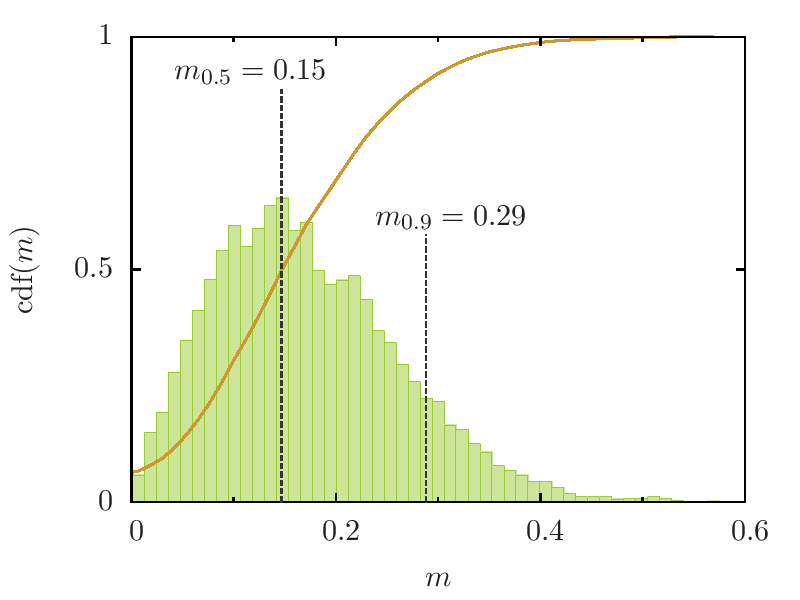}
    \caption{Test of the \EAH{} stochastic template bank for the PMPS
      analysis. The green bars show a histogram of the mismatch
      distribution for 10\,920 noise-free signals from simulated
      pulsars in circular orbits. The orange curve shows the
      cumulative distribution function (cdf) of the mismatch. The
      median $m_{0.5}$ and the 90\%-quantile of the mismatch
      distribution $m_{0.9}$ are highlighted. The template bank covers
      90\% of the parameter space with mismatch $m<0.29$.}
    \or
    \includegraphics[width=\columnwidth]{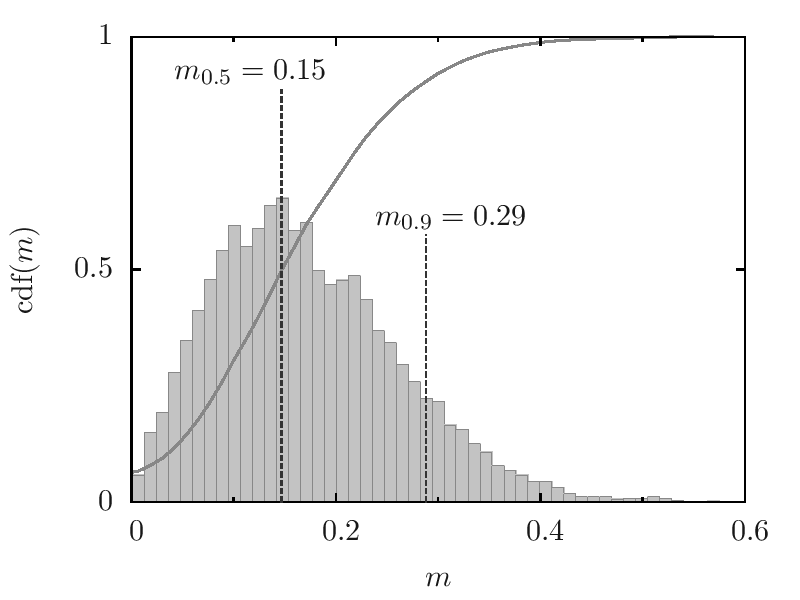}
    \caption{Test of the \EAH{} stochastic template bank for the PMPS
      analysis. The bars show a histogram of the mismatch distribution
      for 10\,920 noise-free signals from simulated pulsars in
      circular orbits. The thick line shows the cumulative
      distribution function (cdf) of the mismatch. The median
      $m_{0.5}$ and the 90\%-quantile of the mismatch distribution
      $m_{0.9}$ are highlighted. The template bank covers 90\% of the
      parameter space with mismatch $m<0.29$}
    \fi
    \label{fig:mismatchhist}
  \end{center}
\end{figure}

The resulting stochastic template bank has a nominal mismatch
$m_0=0.29$ and coverage $\eta = 90\%$.  (The coverage is the fraction
of points in parameter space whose mismatch $m<m_0$ from some
template.)  This efficient template bank contains 12\,140~orbital
templates, c.f.\ about 60~acceleration trials in \citet{ralphpaper}.
The bank was tested using fake noise-free signals at grid points in
orbital parameter space, with $f=f_\text{max}$, where $f_\text{max}$
is the highest spin frequency for which the orbital template bank
achieves the required nominal mismatch, see Sec.~\ref{subsec:searchspace}.
As can be seen from Fig.~\ref{fig:mismatchhist}, the median mismatch
$m_\text{med}=0.15$ is much smaller than the nominal mismatch.

\subsection{Searched Parameter Space}\label{subsec:searchspace}
To conduct a blind search for new pulsars, we must decide what region
of parameter space to cover with a template bank. With unlimited
computing power, we could search any parameter space, no matter how
large.  In practice, the finite computing power of \EAH{} dictates
that we only search some part of the parameter space.

Our choice is motivated first by astrophysical reasons: to target
\EAH{} to an interesting range of putative pulsar spin frequencies and
orbital parameters. Second, for practical reasons we decide to
complete the \EAH{} search on PMPS data within about a year, searching
a previously unexplored part of the orbital parameter space. Thus we
constrain the search parameter space by setting a probabilistic limit
on projected orbital radii, and by an upper limit on spin frequencies.

We estimated the available computing resources based on small-scale
tests on a single computer and on the Atlas computing cluster at the
AEI. The total \EAH{} computing power was estimated from previous 
searches for radio pulsars.

The required number of orbital templates grows $\propto f_\text{max}^3$.
We constrained our orbital template bank to $f_\text{max}=130$\,Hz
($P_\text{min}\approx7.7$\,ms).

Standard acceleration searches lose sensitivity where $P_\text{orb}
\lesssim 10T$.  As discussed in Sec.~\ref{sec:earliersearches},
previous searches \citep{ralphpaper} were
sensitive for orbital periods $P_\text{orb}\lesssim3$\,hrs. Thus we
searched for orbital periods in the range $86\text{\,min}\leq
P_\text{orb} \leq 317\text{\,min}$. The upper boundary was chosen to
provide seamless connection to the sensitivity ranges covered by
previous searches. The lower boundary was dictated by the available
computing power. Sensitivity to orbital periods as short as 86\,min
significantly increases the sensitivity to pulsars in compact binaries. We
added a single template corresponding to an isolated pulsar to sustain
sensitivity to isolated pulsars, or those in very wide orbits.

To provide full sensitivity along the complete orbit of any putative
binary pulsar, the initial orbital phase $\psi$ was not constrained,
covering the range $0\leq \psi < 2\pi$.

We constrained the projected orbital radius by using a probabilistic
bound on the orbital inclination angles based on the masses of
putative pulsars and companions. From Kepler's third law we define
\beq 0 \leq a\sin(i) \leq \alpha \frac{G^\frac{1}{3}
  \Omega_\text{orb}^{-\frac{2}{3}}m_\text{c,max}}{c(m_\text{p,min}
  +m_\text{c,max})^\frac{2}{3}},\label{eq:asinimax} \eeq where
$m_\text{c,max}$ is the maximum companion mass, $m_\text{p,min}$ is
the minimum pulsar mass, $G$ is the gravitational constant, and $c$ is the
speed of light. The constant $\alpha$ measures the probabilistic bound
on orbital inclination angles. For our search, we chose $\alpha = 0.5$,
$m_\text{p,min} = 1.2$\,M$_\odot$ and $m_\text{c,max} =
1.6$\,M$_\odot$. The scaling of the total number of templates with
$\alpha$, the pulsar and companion masses and the orbital period range
is non-trivial.

For different pulsar and companion masses, Eq.~\eqref{eq:asinimax}
defines an upper limit on projected orbital radii as a function of the
orbital angular velocity. For large companion masses, only a fraction
of all possible orbital inclinations fulfill \eqref{eq:asinimax}. In
other words: for smaller companion masses, the condition
\eqref{eq:asinimax} is fulfilled for any $i$, and all of such binary
pulsar systems are detectable by \EAH{}.

\begin{figure}
  \begin{center}
    \ifcase \bwswitch
    \includegraphics[width=\columnwidth]{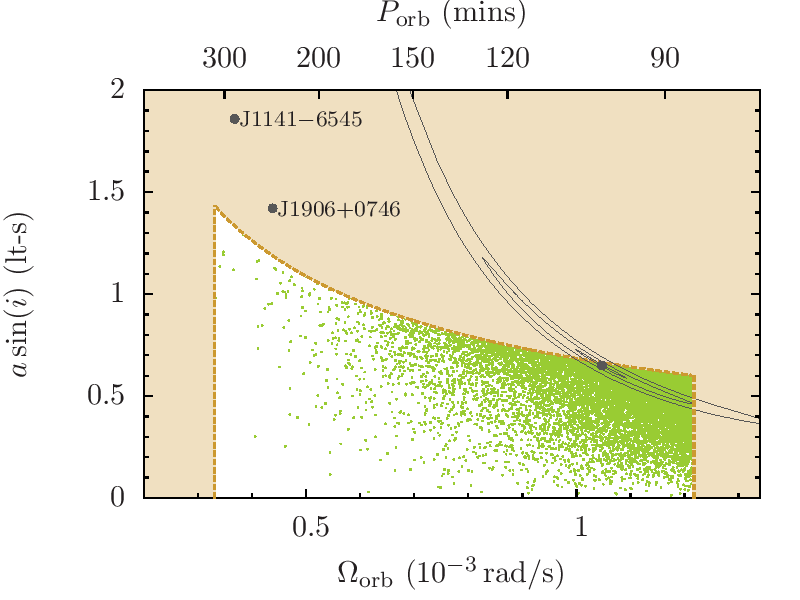}
    \caption{The positions of the orbital templates (green dots) in
      the stochastic template bank constructed for this search. The
      parameter space we searched is shown in white, limited by the
      orange dashed line. The templates shown here are constructed at
      $f_\text{max} = 130$\,Hz. Dark-gray lines show cuts at $\psi=0$
      through surfaces of constant mismatch $m=0.3$ around a single
      template (dark-gray dot) at orbital parameters $P_\text{orb} =
      100$\,min, $a\sin(i) = 0.65$\,lt-s, and $\psi=0$. We chose
      signals at spin frequencies $f$ of 130\,Hz, 65\,Hz, and 15\,Hz,
      respectively. Note that at lower frequencies the enclosed area
      extends far beyond the parameter space covered by the templates,
      enabling discoveries outside the parameter space. Known binaries
      are shown as labeled dark-gray dots.}
    \or
    \includegraphics[width=\columnwidth]{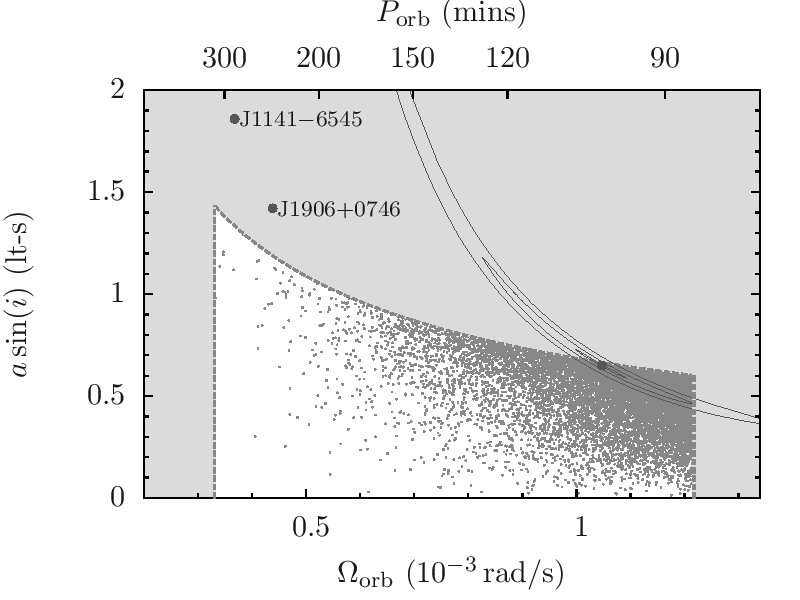}
    \caption{The positions of the orbital templates (dots) in
      the stochastic template bank constructed for this search. The
      parameter space we searched is shown in white, limited by the
      dashed line. The templates shown here are constructed at
      $f_\text{max} = 130$\,Hz. Dark-gray lines show cuts at
      $\psi=0$ through surfaces of constant mismatch $m=0.3$ around a
      single template (dark-gray dot) at orbital parameters
      $P_\text{orb} = 100$\,min, $a\sin(i) = 0.65$\,lt-s, and
      $\psi=0$. We chose signals at spin frequencies $f$ of 130\,Hz,
      65\,Hz, and 15\,Hz, respectively. Note that
      at lower frequencies the enclosed area extends far beyond the
      parameter space covered by the templates, enabling discoveries
      outside the parameter space. Known binaries are shown as labeled
      dark-gray dots.}
    \fi
    \label{fig:templatebank}
  \end{center}
\end{figure}

Fig.~\ref{fig:templatebank} shows the slice of orbital parameter space
defined by these conditions. This choice guarantees that circular
orbits at mass ratios for almost all pulsar-white dwarf binaries and
many double neutron star systems lie inside our search space.

Because we use the same template bank for all spin frequencies, it
`over-covers' the parameter space for frequencies
$f<f_\text{max}$. The required density of templates is highest at $f_\text{max}$, because the
detection statistic depends upon the difference in phase, which varies
most rapidly at the highest frequency. For frequencies higher than
$f_\text{max}$, it `under-covers' the parameter space. In
Fig.~\ref{fig:templatebank}, we show cuts through surfaces of constant
mismatch around a single template with fixed orbital parameters and
varying spin frequency.  At lower frequencies, the region covered by
the templates is ``banana-shaped'' and extends well beyond the
parameter space boundaries defined above. This is why our search
detected the relativistic pulsar J1906$+$0746 (see
Sec.~\ref{sec:postproc}) although its orbital parameters lie outside
the parameter space covered by our template bank.

Our search loses sensitivity to the higher harmonics of fast-spinning
MSPs, and searching for these higher harmonics is
prohibited by the high computing costs. Increasing the value of
$f_\text{max}$ to (say) 1\,kHz, is computationally unfeasible and
would require $\approx 500$ times more computing resources than our
search!

\subsection{Comparison to Standard
  Searching}\label{subsec:compaccsearch}
As discussed in Sec.~\ref{subsec:searchspace}, standard acceleration
searches \citep{2002AJ....124.1788R} lose sensitivity for binary
pulsars with $P_\text{orb}\lesssim10T$. To quantify this effect for the PMPS
data, we numerically computed the mismatch of an acceleration search
and that of the \EAH{} search at different orbital parameter space
points and at the highest spin frequency $f_\text{max}=130$\,Hz. The
orbital parameter space was covered by a cubic grid of 10\,920~points
in $\Omega_\text{orb}$, $a\sin(i)$, and $\psi$.

For each of the orbital parameter space points, we numerically
simulated an acceleration search and the \EAH{} search for a signal at
this point. The simulated acceleration search followed the standard
setup \citep{pulsarhandbook}: We set up a grid of accelerations $a_1$
in a range of $\pm 500$\,m\,s$^{-2}$ and in steps of $\delta
a_1=0.26$\,m\,s$^{-2}$. The acceleration range was chosen to agree with
\citet{ralphthesis}, and the acceleration step size was chosen (very
conservatively) requiring that the signal does not drift by more than
half a Fourier bin as described in Sec.~6.2.1 of \citet{pulsarhandbook}.
Further, we simulated the same frequency resolution $\Delta f = 1/(3T)$
as for the \EAH{} search. Then, we generated noise-free sine-wave signals
given by Eq.~\eqref{eq:fullphase}, and computed the detection statistic
$\mathcal P_1 = S_0$, but using a template phase model $\Phi(t)$ which
is a quadratic function of $t$, corresponding to the signal from a
single-harmonic isolated pulsar moving \textit{with constant
  acceleration}.

For the simulated \EAH{} search, the detection statistic $\mathcal
P_1=S_0$ was computed using a template phase model from
Eq.~\eqref{eq:fullphase}, then the fractional loss of detection
statistic (mismatch $m$) was computed for both the acceleration and
the \EAH{} search.

\begin{figure*}
  \begin{center}
    \ifcase \bwswitch
    \includegraphics[width=\columnwidth]{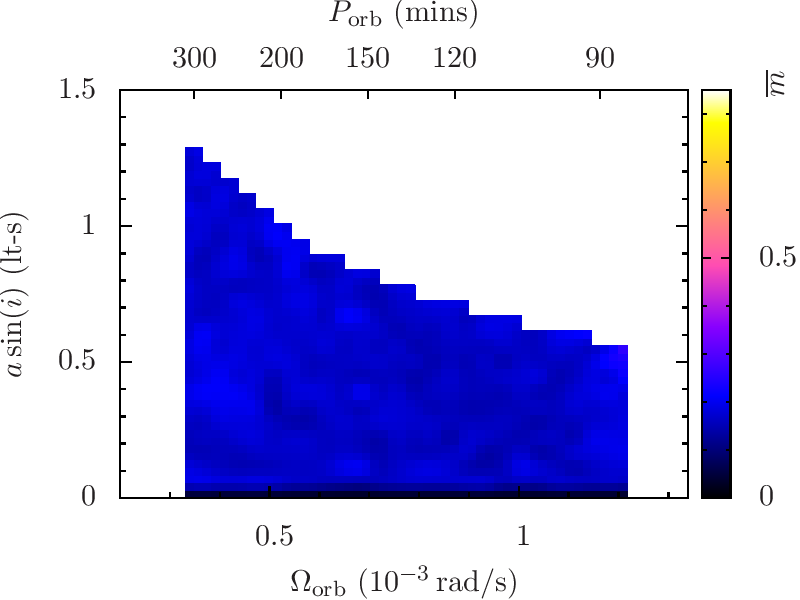}
    \hfill
    \includegraphics[width=\columnwidth]{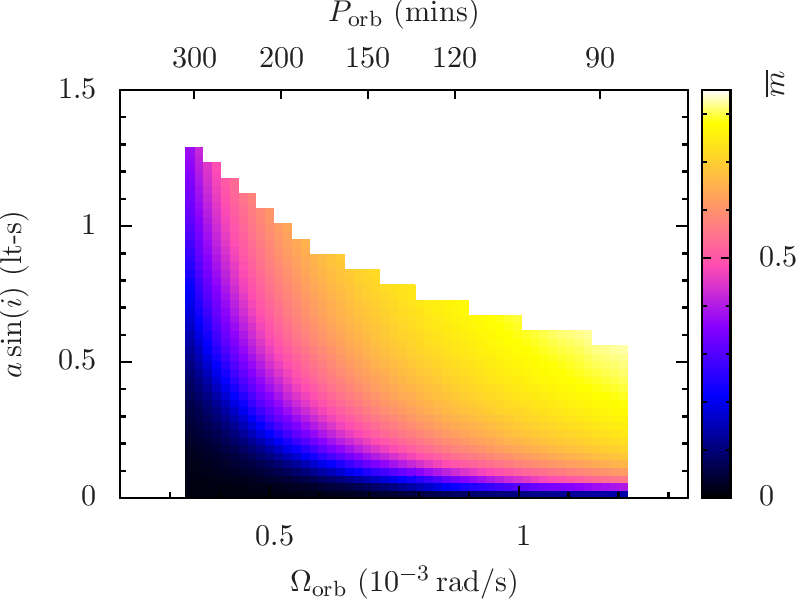}
    \or
    \includegraphics[width=\columnwidth]{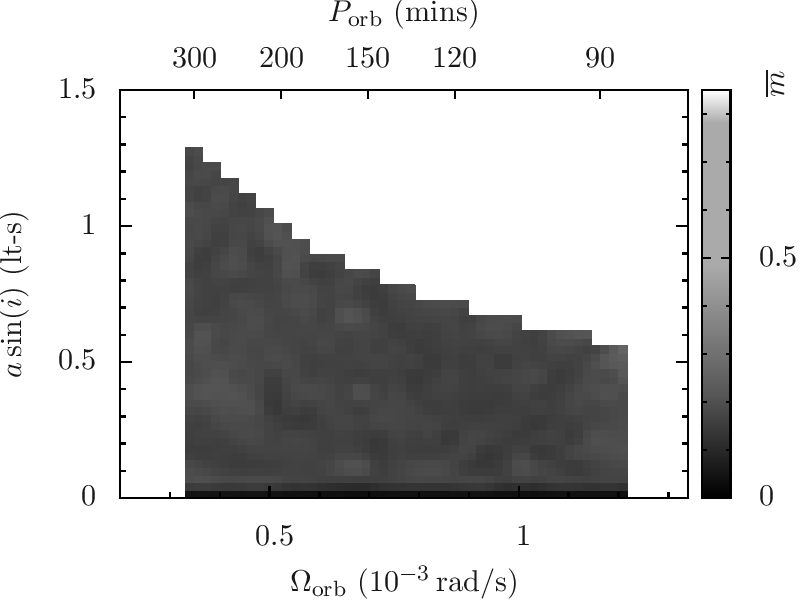}
    \hfill
    \includegraphics[width=\columnwidth]{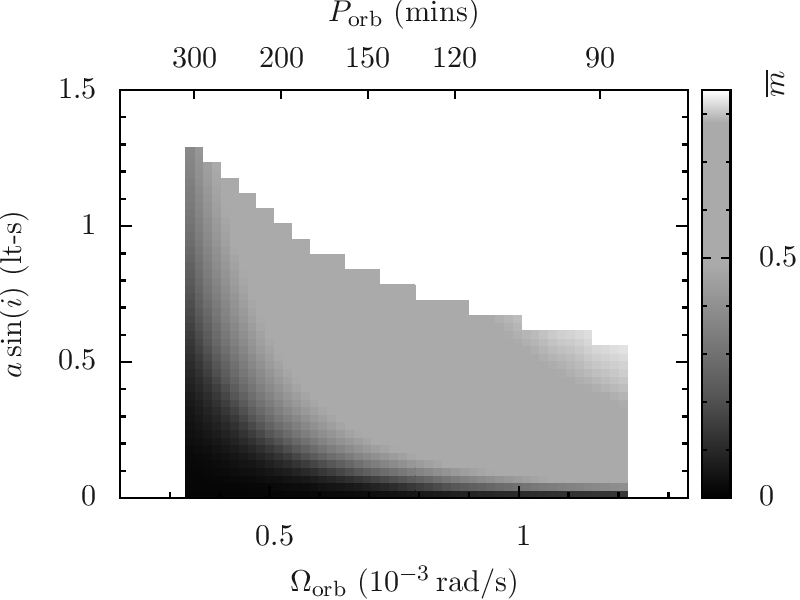}
    \fi
    \caption{Comparison of the mismatch in the \EAH{} processing
      (left) and an acceleration search (right). \textbf{(Left)} The
      mismatch $\overline{m}$, averaged over $\psi$, computed on
      regularly spaced points in the space of $\Omega_\text{orb}$ and
      $a\sin(i)$. The grid consisted of 10\,920 points, in each of
      which a sine-wave noise-free signal was generated and
      searched for with the complete stochastic template bank. The
      coverage is not uniform, because the stochastic template bank is
      based on random template banks. \textbf{(Right)} The results of
      a simulated acceleration search over the same parameter space
      grid. Note the significant loss of sensitivity of the
      acceleration search for higher values of $\Omega_\text{orb}$ and
      $a\sin(i)$.}
    \label{fig:templatebankaccsearch}
  \end{center}
\end{figure*}

Fig.~\ref{fig:templatebankaccsearch} shows the results of our
comparison, where we have averaged the mismatch over the orbital phase
$\psi$. For the simulated \EAH{} search (left panel), the averaged
mismatch $\overline m$ is close to the target value of nominal
mismatch ($m_0 = 0.3$) in the entire parameter space. Because of the
underlying random template bank, there are small parts with slightly
higher and lower mismatches. Close to $a\sin(i)=0$, the mismatch is
considerably smaller, since this parameter space region is covered by
the single template for isolated pulsars.

For the simulated acceleration search (right panel of
Fig.~\ref{fig:templatebankaccsearch}), the mismatch reaches
unacceptably high values $\overline m \gtrsim 0.5$ over a large
fraction of the parameter space. Only in a small part does the
acceleration search achieve mismatches comparable to the our method.
Our results clearly show the improvement in the detection of radio
pulsars in compact binary orbits. Regions of signal parameter space
that were virtually inaccessible with acceleration searches are in
reach of our method.

\subsection{Analysis on the \EAH{} Host Machines}
\label{subsec:analysis}
This section summarizes the ``signal analysis'' part of the
\EAH{} radio pulsar search pipeline, which runs on the
volunteers' hosts and does the bulk of the computing work.
The code is distributed under the GPL~2.0 license and is available
for CPUs and GPUs under Linux, Windows, and Mac OS~X; a complete
description may be found in \citet{2013arXiv1303.0028A}. 

For each host, the input data are typically four de-dispersed time series
which are analyzed sequentially as described here. The search code computes
the detection statistics $S_0, \dots, S_4$ at each template-bank point
in parameter space, and then returns back to the \EAH{} server a list
of ``top candidates'': the points in parameter space where the
detection statistic was largest. The analysis consists of a data
preparation step, a loop over orbital templates, and an output/candidate
reduction step.

In the data preparation step, the input time series is uncompressed
and converted into single-precision floating-point format. It is
whitened in the frequency domain using a running-median average
spectrum. RFI is replaced by computer-generated Gaussian noise using
a list of known contaminated frequency bands in the fluctuation power
spectra of the zero-DM de-dispersed PMPS data \citep{2004MNRAS.355..147F};
this ``zapping'' step uses same list globally for the analysis of
all PMPS beams. After this whitening and cleaning, the data are returned
to the time-domain.

Then the search code begins to iterate through the orbital
templates. For each orbital template, the detection statistics
$S_0,\dots,S_4$. of Eq.~\eqref{eq:statdef} are computed on the full
frequency grid with spacing $\Delta f = 1/3T$. This is done by
resampling the data in the time-domain to remove the effects of the
orbital modulation.  The data are then mean-padded to length $3T$ and
Fourier transformed using a fast Fourier transform (FFT)\footnote{Mean-padding increases the
frequency resolution and avoids loss of detection statistic for signal
frequencies not at the center of the initial Fourier bins.}; the detection
statistic $\mathcal P_n$ is the squared modulus of the Fourier
amplitude of the resampled time series in the $n$th frequency bin.

Inside the loop, the client search code maintains five different lists
of candidates corresponding to the detection
statistics $S_0,\dots,S_4$. Each contains the 100~candidates with the
largest values of $S_i$ having \textit{distinct} values of fundamental
frequency $f$.  The code checkpoints after each loop iteration; if
needed this permits it to be restarted with little loss of compute
time.

When the loop over orbital templates is complete, the statistical
significance $\mathcal S$ is computed for all 500~candidates, and the
100 with the highest $\mathcal S$ are returned to the \EAH{} server in
a single \textit{result file}. Based on false-alarm statistics one can
show that selecting the 100 most significant candidates results in
digging down well into the noise-dominated regime.

The result file is uploaded to the central \EAH{} servers and validated
in a two-step process, described in \citet{2013arXiv1303.0028A}. 
In order to be accepted it has to agree with the result calculated by
another volunteer's computer with a fractional accuracy of about one
part in $10^5$.

\subsection{Manual Candidate Selection}\label{sec:postproc}
After the upload of all 440~result files for each beam, one has to
filter out the most promising candidates from all 44\,000 candidates
in each beam. The majority of candidates in any given beam are caused
by random noise fluctuations and have low significance values
$\mathcal S$. Thus, thresholding on $\mathcal S$ is a possible step in
reducing the number of candidates. However even in the presence of a
highly-significant signal, correlations can cause the signal to show
up at multiple DMs, frequencies, and/or orbital parameters. Thus, more
sophisticated methods are required to reduce the number of candidates
to follow up.
  
\begin{figure}
  \begin{center}
    \ifcase \bwswitch
    \includegraphics[width=\columnwidth]{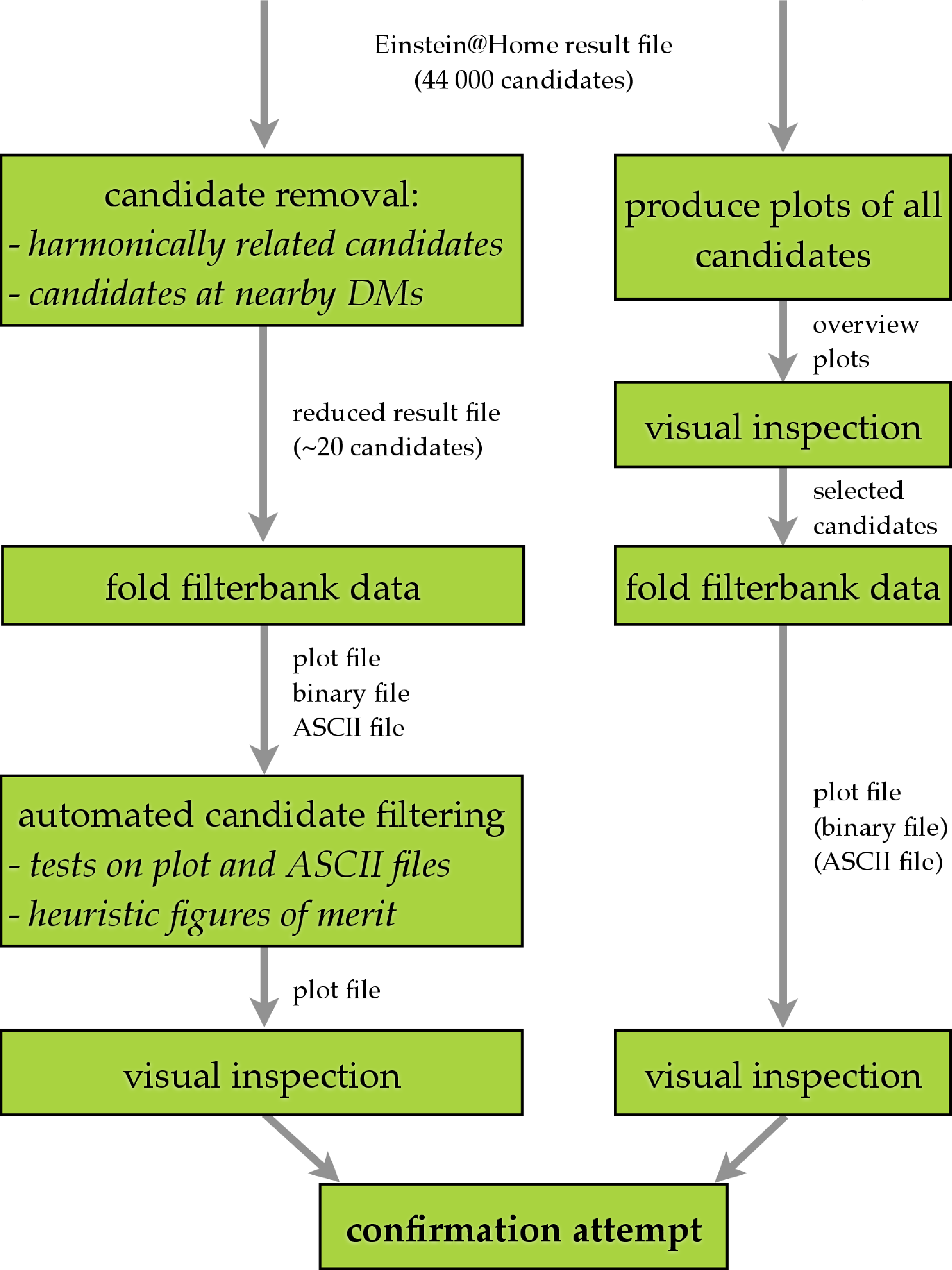}
    \or
    \includegraphics[width=\columnwidth]{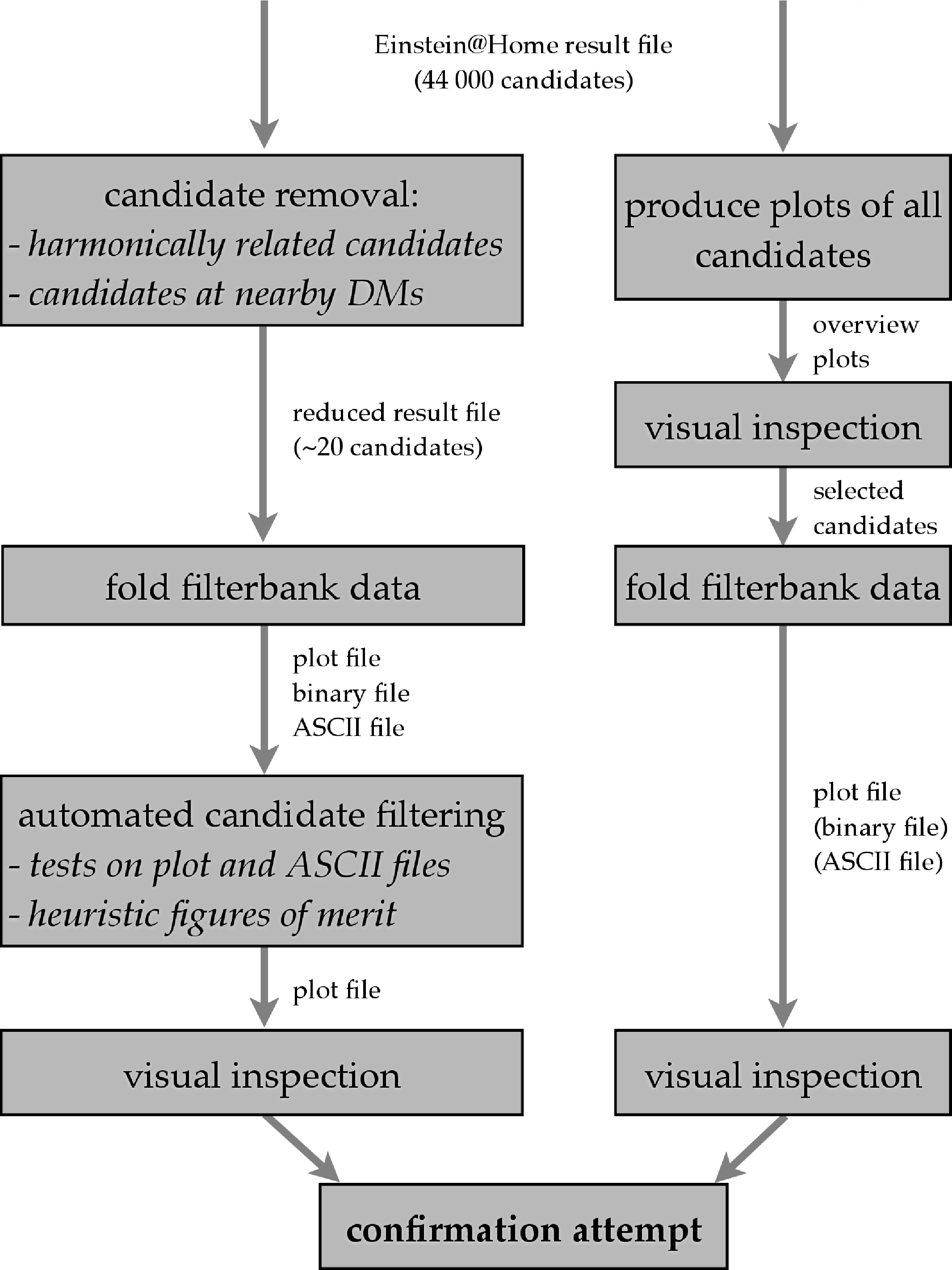}
    \fi
    \caption{Turning the \EAH{} results into confirmations of new
      pulsars: this diagram shows the steps of the two post-processing
      methods used in our search. The right-hand side displays the
      methods using the visual inspection of overview plots described
      in Sec.~\ref{sec:postproc}. The left-hand side shows the steps
      of the automated post-processing methods to reduce the number of
      candidates from Sec.~\ref{subsec:autopostproc}.}
    \label{fig:postproc}
  \end{center}
\end{figure}

We used two methods to identify pulsar-like signals, employing
frequency coordinates tailored to the binary parameter space. One method is
based on producing overview plots that allow a quick and easy
identification of pulsars and first estimates of their orbital
parameters. The second method employs automated filtering algorithms
to identify promising candidates in the binary parameter
space. Fig.~\ref{fig:postproc} shows a flow diagram comparing both
methods, which we describe in the following.

Our first method for the identification of promising pulsar candidates
in the \EAH{} result files uses custom-made overview plots. These
visualize the complete set of candidates for any given PMPS
observation. We identified pulsars by characteristic patterns in these
plots. Promising candidates are followed up with tools from the
\textsc{presto} software suite.

The set of overview plots is automatically produced for visual
inspection when all valid result files for a given beam are available
on the \EAH{} servers.  The plots show the significance for all
44\,000 candidates as a function of a the spin frequency at the
detector $\nu_1$, the associated spin frequency derivative $\nu_2$,
and combinations of the orbital template parameters. The
coordinates $\nu_1$ and $\nu_2$ resolve some of the correlations of
the detection statistic $\mathcal P_n$ in the orbital parameters. The
$\nu_1$ and $\nu_2$ are the linear and quadratic coefficients in a
polynomial expansion of the phase model \eqref{eq:fullphase} in
$t$. Their derivation may be found in the Appendix.

\begin{figure*}
  \begin{center}
    \ifcase \bwswitch
    \includegraphics[width=\columnwidth]{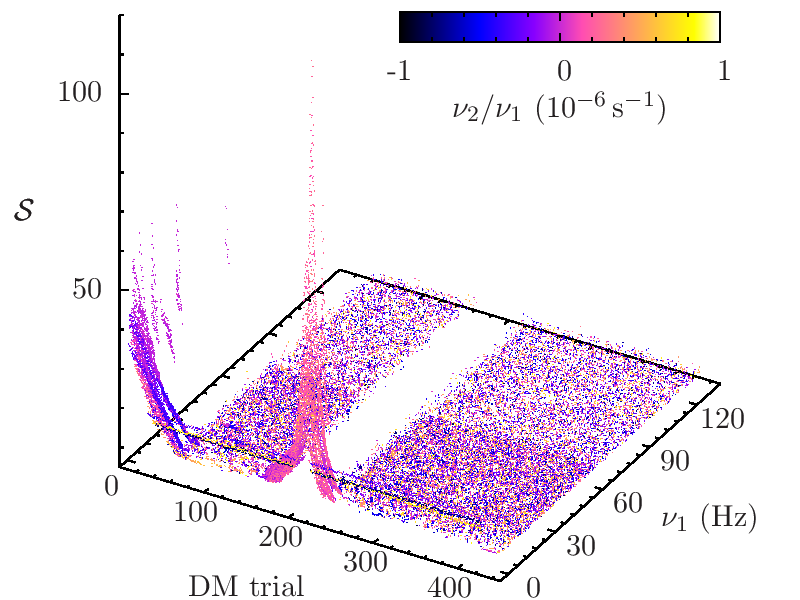}
    \hfill
    \includegraphics[width=\columnwidth]{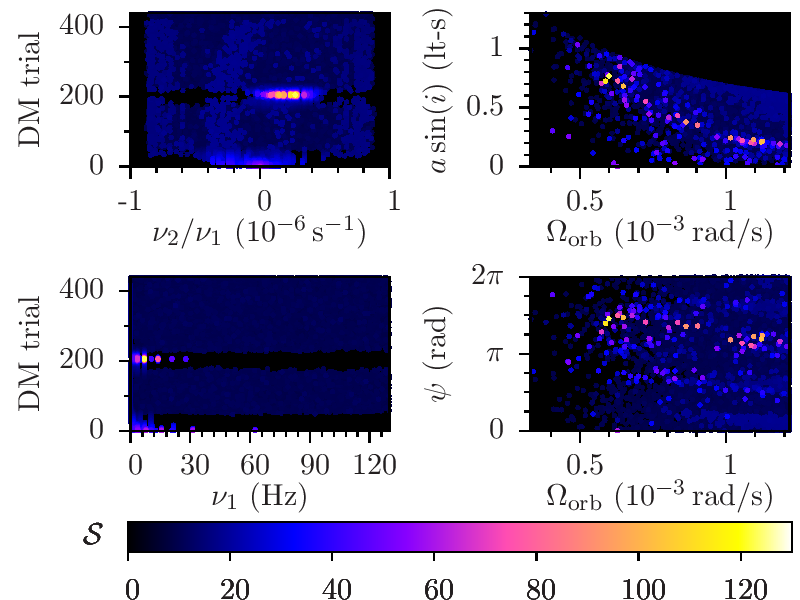}
    \or
    \includegraphics[width=\columnwidth]{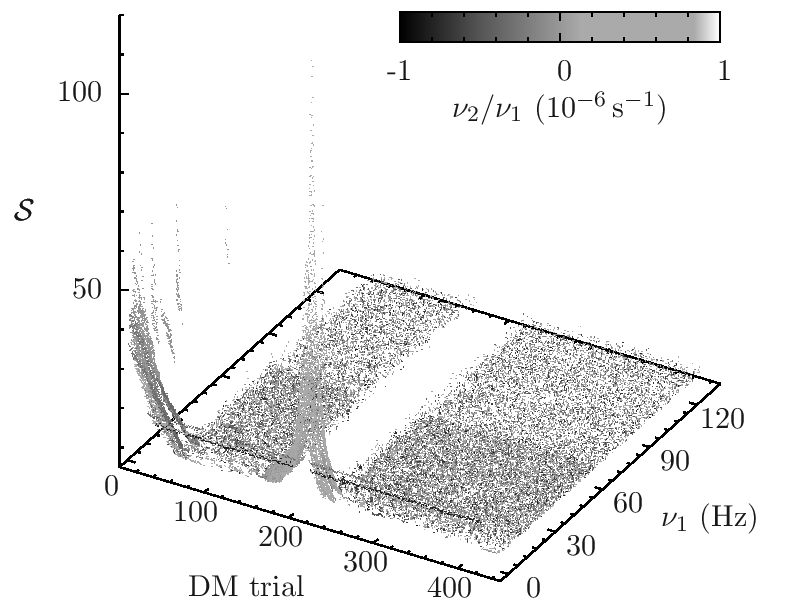}
    \hfill
    \includegraphics[width=\columnwidth]{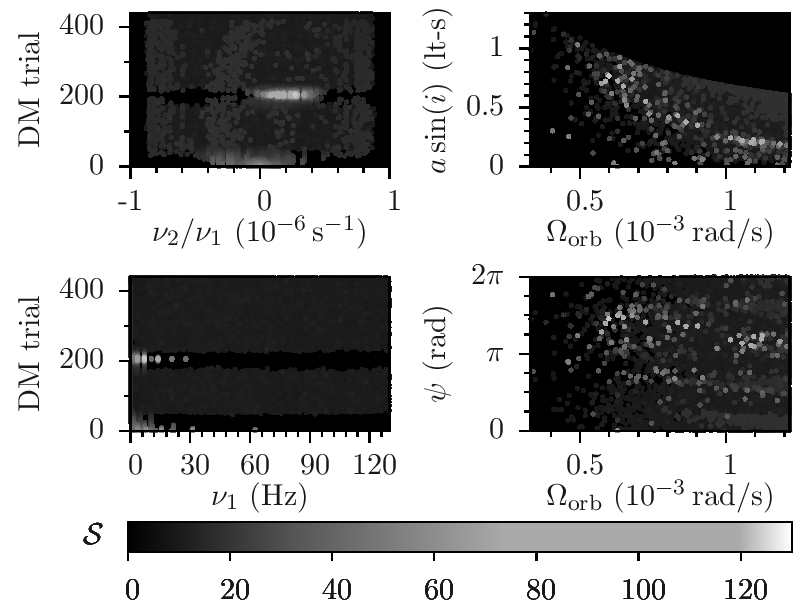}
    \fi
    \caption{Example post-processing overview plots, showing the
      highly-significant detection of the binary pulsar
      J1906$+$0746. \textbf{(Left)} The plot shows the significance
      $\mathcal S$ as a function of the DM trial number and the spin
      frequency $\nu_1$ at the detector of each candidate. The color
      code displays the relative change in spin frequency
      $\nu_2/\nu_1$ from orbital motion.  The coordinates $\nu_1$ and
      $\nu_2$ are the first two coefficients of the polynomial
      expansion of the orbital phase as given in the
      Appendix. Since the top 100~candidates
      are reported for each DM trial and the pulsar is detected with
      very high significance, there are no detections of the noise
      floor in a DM range around the pulsar. The missing noise floor
      at low DM is caused by remnant RFI discernible by increasing
      $\mathcal S$ towards $\text{DM}=0$\,pc\,cm$^{-3}$. \textbf{(Right)} The
      four sub-panels show the significance $\mathcal S$ in color-code
      as a function of different combinations of spin frequency and
      the orbital parameters. The top left shows $\mathcal S$
      projected onto the space of relative frequency change
      $\nu_2/\nu_1$ and DM trial number, the top right plot shows
      $\mathcal S$ in the space of orbital angular velocity
      $\Omega_\text{orb} = 2\pi/P_\text{orb}$ and projected orbital
      radius $a\sin(i)$. The bottom left plot displays $\mathcal S$ as
      a function of spin frequency $\nu_1$ and DM trial number, the
      bottom right plot shows $\mathcal S$ in the space of orbital
      angular velocity $\Omega_\text{orb}$ and initial orbital phase
      $\psi$. Note the clearly visible non-zero value of $\nu_2/
      \nu_1$ caused by the orbital motion of the pulsar in its 4-hr
      orbit during the 35\,min observation. Even though only about
      15\% of its orbit was observed, the right-hand side panels
      already allow some first constraints of the orbital parameters.}
    \label{fig:ovplot}
  \end{center}
\end{figure*}

An example of the five different plots in our post-processing is given
in Fig.~\ref{fig:ovplot}, which shows the highly-significant detection
of the binary pulsar J1906$+$0746 in the \EAH{} results.

The number of candidates identified from these overview plots is
relatively small (a few per beam at most, none in the majority of
beams, and on average one candidate in $\sim50$~beams). Tools from the
\textsc{presto} software suite are used to fold the full-resolution
filterbank data at the candidate spin period and DM identified from
the \EAH{} results and to optimize the parameters. In total we
inspected of order 1000 of these folded time series plots by eye
and used them to judge the broadband nature and temporal continuity of
the signal. The ATNF
catalogue\footnote{\url{http://www.atnf.csiro.au/people/pulsar/psrcat/}}
\citep{2005yCat.7245....0M}, and Web sites listing known, but as yet
unpublished, pulsars\footnote{\url{http://astro.phys.wvu.edu/GBTdrift350/},
  \url{http://www.physics.mcgill.ca/~hessels/GBT350/gbt350.html},
  \url{http://astro.phys.wvu.edu/dmb/},
  \url{http://www.naic.edu/~palfa/newpulsars/}} are checked to ensure
that the candidate is not a detection of a known pulsar. About half
of our new discoveries were found by the visual inspection of overview
plots.

\subsection{Automated Candidate Selection}\label{subsec:autopostproc}
The visual inspection of overview plots is augmented by an automated
post-processing stage. This sifts through all candidates in a given
beam and identifies the most promising ones for a follow up via
folding of the filterbank data. With infinite computing and man power,
all candidates of a given beam would be used for folding the
filterbank data.  In practice, this is neither feasible nor
necessary. Folding the 44\,000 candidates for any given beam
requires an unfeasibly large amount of computing time and human
resources for further inspection. Many of the candidates will be caused either from random noise
or by a \textit{single} pulsar (or RFI) signal. For example, following
up candidates at harmonically related frequencies and nearby DM trials
does not yield additional information (c.f.\ Fig.~\ref{fig:ovplot}).

The first goal of the automated post-processing is to reduce the
number of candidates from the complete results files. The second goal
is to find those pulsars whose identification with the overview plots
is difficult. Weak signals at higher spin frequencies may only be
registered at a few trial DMs above the noise level, making it
difficult to detect them visually.

The automated post-processing consists of three steps.
In the first step we reduce the number of harmonically related
candidates. We start at the candidate with the highest value of
$\mathcal S$. Candidates associated with sub-harmonic and harmonic
frequencies of this candidate are flagged as possibly related as follows. We
conservatively assume a frequency band around the candidate's value of
$\nu_1$ with a width $\delta \nu_1$ given by the maximum Doppler
modulation of all templates in the orbital template bank. The width of
the frequency band is obtained from maximizing the frequency
modulation amplitude $\Omega_\text{orb}a\sin(i)/c$ over the entire
orbital template bank and computing the Doppler range at frequency
$\nu_1$ from \beq\delta \nu_1 = \nu_1
\max(\Omega_\text{orb}a\sin(i)/c).  \eeq After this step, only
candidates in the Doppler range around the fundamental frequency are
retained for the next step.

These are further winnowed down in the second step. Real pulsar candidates
should produce a peak in significance as a function of DM, separated from
the noise-dominated background. To take this property into account, we
approximate the significance as a function of DM
\citep{2003ApJ...596.1142C} by a parabola \beq \mathcal S(\text{DM}) =
\mathcal S_0 \left(1- \frac{\text{DM} - \text{DM}_0} {\Delta
    \text{DM}}\right)^2.
\label{eq:parabola}
\eeq Here, $\mathcal S_0$ is the significance of the most significant
candidate at central dispersion measure $\text{DM}_0$, and $\Delta
\text{DM}$ is the ``width'' of the parabola, such that
$S(\text{DM}_0\pm\Delta \text{DM})=0$.

Now, we step over a fixed grid of ten trial values of
$\Delta\text{DM}$ between 0.6\,pc\,cm$^{-3}$ and 50\,pc\,cm$^{-3}$ and
determine the best-fit parabola parameter $\Delta\text{DM}$, i.e.\ the
one with the smallest squared residuals.  To make sure that the
parabola approximation is accurate, we select candidates near the top
of the peak with $\mathcal S \geq 0.6\mathcal S_0$.  Out of all
candidates within the Doppler frequency range and the best fitting DM
range we only keep the most significant candidate.

The procedure of removing harmonically related candidates and those at
similar DMs described above is iterated with the next most significant
candidate until only ``independent'' candidates are left.

The number of remaining candidates per beam was about 20 after this
first reduction step.  We then folded all reduced candidates using the
\textsc{presto} software on the Atlas cluster at the AEI. Each
candidate was folded at fixed DM with five different spin frequencies
between $\nu_1 - 2\delta\nu$ and $\nu_1 + 2\delta\nu$. Here, $\delta \nu$ is the
expected Doppler shift in spin frequency caused by orbital motion and
is computed from the orbital parameters of the candidate. Folding at a
conservatively wide parameter range accommodates possible offsets from
the true orbital parameters. In this step, we folded of order 4\,000\,000
candidates for the whole PMPS search. Due to oversight, we folded the
original data, not the ones resampled at the orbital parameters. We will
re-run our post-processing and report results in a future paper.

The final step of our automated post-processing is done by an algorithm
implemented in \textsc{idl}\footnote{\url{www.exelisvis.com/idl/}}, which
selects pulsar-like candidates based on figures of merit derived from
the \textsc{prepfold} plots and associated binary and ASCII files.

The algorithm first checks for disqualifying metrics in the ASCII file
associated with the folded data output. Features that cause rejection
include periods that fall within the range of known interference
signals and a signal strength below a chosen threshold (5.0$\sigma$ as
computed by \textsc{prepfold}).

For output files that survive these tests, the software uses two
different tests applied directly to the plots produced by
\textsc{prepfold}. It applies them separately to two parts of each
plot, see e.g.\ Fig.~\ref{fig:discoveries}. The first test looks for
signal strength and consistency in phase in the time versus phase plot
(left hand bottom subplot in Fig.~\ref{fig:discoveries}). A pulsar-like
signal is a straight, vertical line when folded with the proper
parameters. An average intensity is computed for each vertical line at
every horizontal phase position by summing the pixel values in that
column and dividing by the number of pixels summed. A pixel smoothing
of 3, 9, and 21 phase bins is applied in the phase (horizontal)
direction prior to taking each sum. This smoothing accounts for
different possible pulse widths. The resulting average intensity values
for the vertical columns are then compared to with an off-pulse average
intensity value, and a different threshold test is applied for each
smoothing case.

A second test looks at a contour plot showing signal strength as a
function of trial period and period derivative (bottom right sub-plot
in Fig~\ref{fig:discoveries}). The expected signal is one that is
localized with one significant peak. The algorithm counts and
classifies the number of consolidated strong response regions
(determined by the number of contiguous pixels that are saturated)
into ``large'' and ``small'' areas, which have sizes of at least 8 and
250 pixels, respectively. These correspond to 0.03\% and 0.86\% of
the total area of the bottom right sub-plot in
Fig~\ref{fig:discoveries}.  The code retains the candidate as valid if
one or two large areas and no more than 13 total large and small
areas are counted, or if there are less than eight total large and small
areas (regardless of the number of large areas).  Violation of either
of these conditions indicates that the plot is not likely to reflect a
unique period and period derivative combination. Testing for both
small and large areas was chosen to account for the fact that not all
pulsar signals will be strong enough to yield just a single contour
peak.

The thresholds described above were chosen by using a subset of the
PALFA data collected with the WAPP backends \citep{2006ApJ...637..446C}
as a test case. A variety of area sizes and threshold numbers were
tried for the tests to optimize the results (that is, until the number
of candidates passing the test was reduced as much as possible without
eliminating any known pulsars).

The \textsc{idl} code reduces the number of candidates by up to a
factor of a hundred, leaving a number ($\sim100\,000$) that we individually
checked by eye.

Our two different pulsar identification methods have different
selection effects, making them sensitive to different classes of
pulsars. The automated processing has proven very useful in
identifying signals which were missed in the overview plots, because
they were weak and/or masked by RFI (cf.\ Tab.~\ref{tab:all}). But, since
visual inspection of result plots and human judgment is the final step
in both methods, human error is a possible source for the rejection of
real pulsar signals. The simplified folding at five different spin
frequencies as described above can lower the sensitivity to the most
extreme compact binary pulsars in the automated candidate selection.

\section{\EAH{} Discoveries in the PMPS
  Data}\label{sec:discoveries}
Any pulsar candidate surviving the checks described above was
scheduled for re-observation to confirm the celestial nature of the
candidate signal. We confirmed the pulsar discoveries presented here
using the Parkes telescope, the Lovell telescope at Jodrell Bank, and
the Effelsberg telescope. After the initial discovery, regular timing
observations were conducted to further characterize the pulsar and a
possible binary system. All timing solutions presented in this
publication were obtained with observations at the Lovell telescope.

In total, 24 previously unknown pulsars have been identified. Most of
the new discoveries are relatively faint, with period-averaged flux
densities between 0.1\,mJy and 2.7\,mJy at 1.4\,GHz. The spin periods
of the discoveries lie between 3.78\,ms and 2624\,ms. Eighteen pulsars
are isolated, and six are members of binary systems.

Follow-up observations at Jodrell Bank used a dual-polarization
cryogenic receiver on the 76-m Lovell telescope, having a system
equivalent flux density of 25\,Jy. Data were processed by a
digital filterbank which covered the frequency band between 1350\,MHz
and 1700\,MHz with channels of 0.5\,MHz bandwidth. We typically made
observations with a total duration of between 10\,min and 40\,
min, depending upon the discovery signal-to-noise ratio.  Data
were folded at the nominal topocentric period of the pulsar for
sub-integration times of 10\,s.  After inspection and `cleaning'
of any RFI, we de-dispersed profiles at the nominal value of the
pulsar DM. Initial pulsar parameters were established by conducting
local searches in period and DM about the nominal discovery values and
finally summed over frequency and time to produce integrated
profiles. Time of arrival (TOA) data were obtained after matching with
a standard template and processed using standard analysis techniques
with \textsc{psrtime} and \textsc{tempo}.

In the following, we present all 24~discoveries. Tab.~\ref{tab:all}
shows the properties of all new pulsars from our search. For pulsars
without a coherent timing solution, the sky position was derived from
the discovery beam center coordinates or weighted averages in case of
detections in multiple beams. The position error is given by the PMPS
beam size. The spin period $P$ for pulsars without coherent timing
solution is from the discovery observation. We calculated the flux
densities $S_{1400}$ from the discovery observations using the
radiometer equation, e.g.\ \citep{pulsarhandbook} with the gain and system
temperature of the Parkes 21-cm Multibeam Receiver \citep{1996PASA...13..243S}. The DM is the
nominal value from the discovery observation. The distance $D$
was estimated based on a Galactic electron density model from
\citet{2002astro.ph..7156C} with typical errors at the level of
$\sim$20\%. For reproducibility of our results, we quote the discovery
PMPS beam and the \EAH{} significance $\mathcal S$ from
Eq.~\eqref{eq:sig}. The last column shows the discovery method, either
`V' for the visual inspection of overview plots, or `A' for the
automated post-processing. A total of 13~pulsars were found by the
first method, the remaining 11 by the second method. We have found
coherent timing solutions for five of our discoveries, marked in the
table with $^\dagger$.

\begin{table*}
  \caption{\rm Pulsars discovered by the \EAH{} analysis of the
    Parkes Multi-beam Pulsar Survey data. Sources with a fully-determined
    timing solution are marked with $^\dagger$ and pulsars in binary systems
    by $^\flat$. For these sources, the values in parentheses are the 1-sigma
    errors as reported by \textsc{tempo}. For pulsars without a coherent timing
    solution, right ascension (RA) and declination (Dec) are the beam center
    coordinates  or weighted averages in case of detections in multiple beams.
    Their position error is estimated from the PMPS beam size. Their spin periods
    $P$ and dispersion measure DM are the values in the discovery observations.
    $S_{1400}$ denotes flux densities and $D$ the estimated distance. The discovery
    PMPS beam and the \EAH{} significance $\mathcal S$ are given for reproducibility
    reasons. The last column shows the discovery method: `V' for the visual
    inspection, or `A' for the automated method.}
  \setlength{\extrarowheight}{2pt}
  \label{tab:all}
  \begin{tabular*}{\textwidth}{@{\extracolsep{\fill}}llllllrrrlrr}
    \hline
    \multicolumn{1}{c}{PSR} & \multicolumn{1}{c}{RA} & \multicolumn{1}{c}{Dec} & \multicolumn{1}{c}{$P$} & \multicolumn{1}{c}{$\dot P$} & \multicolumn{1}{c}{$P$ Epoch} & \multicolumn{1}{c}{DM} & \multicolumn{1}{c}{$S_{1400}$} & \multicolumn{1}{c}{$D$} & \multicolumn{1}{c}{PMPS beam} & \multicolumn{1}{c}{$\mathcal S$} &\\
    & \multicolumn{1}{c}{(J2000)} & \multicolumn{1}{c}{(J2000)} & \multicolumn{1}{c}{(s)} & \multicolumn{1}{c}{$\left(10^{-15}\right)$} & \multicolumn{1}{c}{(MJD)} & \multicolumn{1}{c}{(pc\,cm$^{-3}$)} & \multicolumn{1}{c}{(mJy)} & \multicolumn{1}{c}{(kpc)} & &\\
    \hline
    J0811$-$38 & 08:11.7(5) & $-$38:57(7) & 0.482594(2) & -- & 50824.5 & 336.2 & 0.3 & 6.2 & 0026\_0051 & 15.6 & V\\

    J1227$-$6208$^\flat$ & 12:27.6(5) & $-$62:10(7) & 0.034529685(8) & -- & 51034.1 & 363.2  & 0.8 & 8.4 & 0058\_036D & 17.9 & A\\

    J1305$-$66 & 13:05.6(5) & $-$66:39(7) & 0.1972763(2) & -- & 51559.7 & 316.1  & 0.2 & 7.5 & 0109\_0033 & 15.5 & A\\

    J1322$-$62 & 13:22.9(5) & $-$62:51(7) & 1.044851(4) & -- & 50591.6 & 733.6  & 0.3 & 13.2 & 0001\_0016 & 23.1 & V\\

    J1455$-$59 & 14:55.1(5) & $-$59:23(7) & 0.1761912(2) & -- & 50841.7 & 498.0  & 1.6 & 7.0 & 0038\_0182 & 14.0 & V\\

    J1601$-$50 & 16:01.4(5) & $-$50:23(7) & 0.860777(4) & -- & 50993.6 & 59.0 & 0.4 & 3.6 & 0042\_0039 & 29.1 & A\\

    J1619$-$42 & 16:19.1(5) & $-$42:02(7) & 1.023152(4) & -- & 51975.6 & 172.0 & 0.6 & 3.7 & 0137\_041B & 35.4 & V\\

    J1626$-$44 & 16:27.0(5) & $-$44:22(7) & 0.3083536(5) & -- & 51718.6 & 269.2 & 0.3& 4.8 & 0125\_077C & 13.2 & A\\

    J1637$-$46 & 16:37.6(5) & $-$46:13(7) & 0.493091(2) & -- & 50842.9 & 660.4 & 0.7 & 7.0 & 0039\_0055 & 17.2 & V\\

    J1644$-$44 & 16:44.6(5) & $-$44:10(7) & 0.1739106(2) & -- & 51030.2 & 535.1 & 0.4 & 6.2 & 0056\_020B & 14.1 & V\\

    J1644$-$46 & 16:44.1(5) & $-$46:26(7) & 0.2509406(1) & -- & 50839.0 & 405.8 & 0.8 & 4.8 & 0035\_0293 & 13.2 & A\\

    J1652$-$48$^\flat$ & 16:52.9(5) & $-$48:45(7) & 0.0037851238(4) & -- & 51373.3 & 187.8 & 2.7 & 3.3 & 0085\_0254 & 22.3 & A\\

    J1726$-$31$^\flat$ & 17:26.6(5) & $-$31:57(7) & 0.12347018(9) & -- & 51026.4 & 264.4 & 0.4 & 4.1 & 0054\_015A & 15.9 & A\\

    J1748$-$3009$^\flat$ & 17:48:23.79(2) & $-$30:09:12.2(5) & 0.009684273(2) & -- & 51495.1 & 420.2 & 1.4 & 5.0 & 0102\_0059 & 18.0 & A\\

    J1750$-$2536$^\flat$ & 17:50:33.39(2) & $-$25:36:43(3) & 0.034749053(8) & -- & 50593.8 & 178.4 & 0.4 & 3.2 & 0002\_0089\footnotemark & 15.9 & A\\

    J1755$-$33 & 17:55.2(5) & $-$33:31(7) & 0.959466(4) & -- & 52080.6 & 266.5 & 0.2 & 5.7 & 0141\_0097\footnotemark & 21.2 & V\\

    J1804$-$28 & 18:04.8(5) & $-$28:07(7) & 1.273011(9) & -- & 51973.7 & 203.5 & 0.4 & 4.2 & 0137\_039B & 13.2 & A\\

    J1811$-$1049$^\dagger$ & 18:11:17.07(8) & $-$10:49:03(4) & 2.6238585620(3) & 0.8(2) & 55983.5 & 253.3 & 0.3 & 5.5 & 0149\_0108 & 29.2 & V\\

    J1817$-$1938$^\dagger$ & 18:17:06.82(8) & $-$19:38.6(2) & 2.0468376289(2) & 0.36(9) & 55991.8 & 519.6 & 0.1 & 8.6 & 0011\_0323\footnotemark & 16.9 & V\\

    J1821$-$0331$^\dagger$ & 18:21:44.70(3) & $-$03:31:12.7(1) & 0.90231562918(4) & 2.53(2) & 55980.9 & 171.5  & 0.2 & 4.3 & 0148\_0197 & 28.3 & V\\

    J1838$-$01 & 18:38.5(5) & $-$01:01(7) & 0.1832948(2) & -- & 51869.1 & 320.4 & 0.3 & 6.9 & 0132\_0627 & 16.7 & V\\

    J1838$-$1849$^\dagger$ & 18:38:33.79(4) & $-$18:49:59(5) & 0.48824200896(3) & 0.04(1) & 55991.9 & 169.9 & 0.4 & 4.5 & 0140\_0064 & 31.7 & V\\

    J1840$-$0643$^{\flat\dagger}$ & 18:40:09.44(5) & $-$06:43:47(1) & 0.0355778755(3) & 0.2202(7) & 55930.0 & 500.0 & 1.2 & 6.8 & 0060\_0206\footnotemark & 18.2 & V\\

    J1858$-$0736 & 18:58:44.3(7) & $-$07:37(7) & 0.551058591(2) & 5.06(7) & 56108.5 & 194.0  & 0.3 & 5.0 & 0143\_0051 & 16.7 & A\\
    \hline
  \end{tabular*}
\footnotetext[1]{J1750$-$2536 was independently detected in the PMPS beam 0002\_0096.}
\footnotetext[2]{J1755$-$33 was independently detected in the PMPS beams 0136\_0268 and 0118\_021A.}
\footnotetext[3]{J1817$-$1938 was independently detected in the PMPS beam 0043\_0014.}
\footnotetext[4]{J1840$-$0643 was independently detected in the PMPS beam 0087\_0026.}
\end{table*}

\begin{figure}
  \begin{center}
    \ifcase \bwswitch
    \includegraphics[width=\columnwidth]{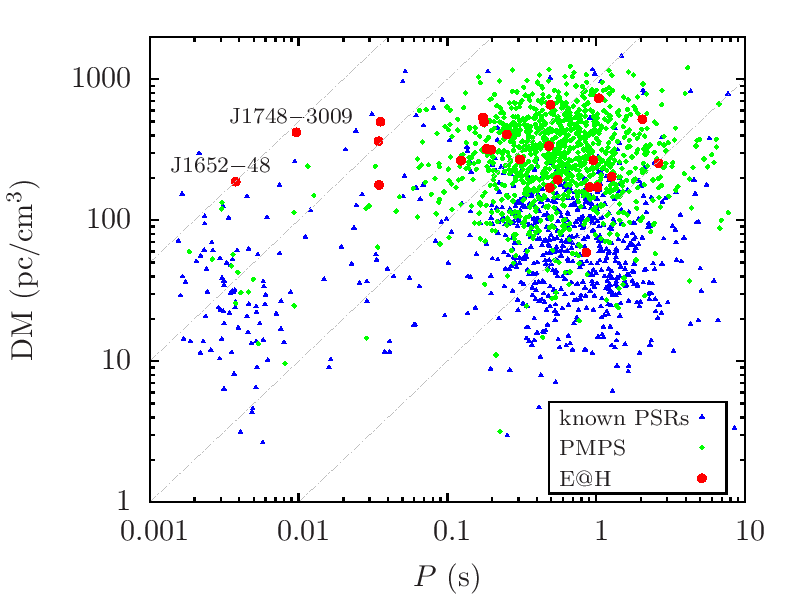}
    \caption{Our discoveries in comparison with the known pulsar
      population in the Galactic field. The plot shows the DM vs.\
      the period for all known pulsars--excluding sources in globular
      cluster and extra-galactic pulsars--from the ATNF catalog
      (blue triangles), discoveries from earlier PMPS analyses (green
      diamonds), and our new discoveries (red circles). The
      distribution of our discoveries at higher DMs is apparent, and
      agrees with the earlier PMPS discoveries. Notable are the
      discoveries of pulsars at lower spin periods and high DMs,
      especially that of PSR~J1748$-$3009, the millisecond pulsar with
      the highest known DM. Dashed gray lines show points of
      $\text{DM}/P=\text{const}$, from top left to bottom right for
      $\text{DM}/P=50, 10, 1, 0.1$\,pc\,cm$^{-3}$\,ms$^{-1}$.}
    \or
    \includegraphics[width=\columnwidth]{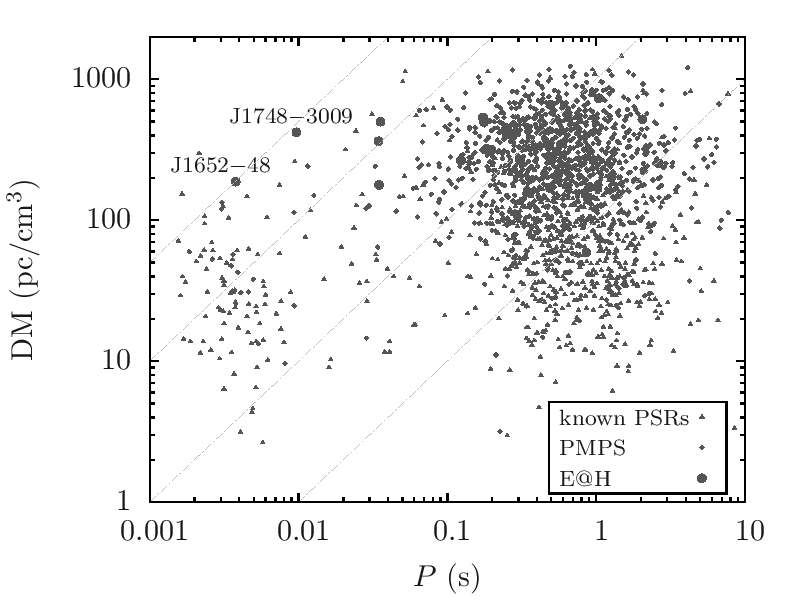}
    \caption{Our discoveries in comparison with the known pulsar
      population in the Galactic field. The plot shows the DM vs.\
      the period for all known pulsars--excluding sources in globular
      cluster and extra-galactic pulsars--from the ATNF catalog
      (triangles), discoveries from earlier PMPS analyses (diamonds),
      and our new discoveries (circles). The distribution of our
      discoveries at higher DMs is apparent, and agrees with the
      earlier PMPS discoveries. Notable are the discoveries of pulsars
      at lower spin periods and high DMs, especially that of
      PSR~J1748$-$3009, the millisecond pulsar with the highest known
      DM. Dashed gray lines show points of $\text{DM}/P=\text{const}$,
      from top left to bottom right for $\text{DM}/P=50, 10, 1,
      0.1$\,pc\,cm$^{-3}$\,ms$^{-1}$.}
	\fi
    \label{fig:dmvsp}
  \end{center}
\end{figure}

To compare our discoveries with the known population, we show DM and
spin periods of our discoveries, together with pulsars in the ATNF
catalogue \citep{2005yCat.7245....0M} and those from earlier PMPS
analyses \citep{2001MNRAS.328...17M, 2002MNRAS.335..275M,
  2003MNRAS.342.1299K, 2004MNRAS.352.1439H, 2004MNRAS.355..147F,
  2006MNRAS.372..777L, 2009MNRAS.395..837K, 2010MNRAS.401.1057K,
  2012ApJ...759..127M, ralphpaper} in Fig.~\ref{fig:dmvsp}. Most
of our discoveries are at large DMs relative to the bulk of the
ATNF pulsars, but their DM distribution agrees with that of other
PMPS pulsars.

Notable are the discoveries at high DMs and small spin periods. This
combination of DM and $P$ usually hampers the detection of
pulsars, since the channel smearing increases with larger DMs and with
smaller periods. Longer sampling times further decrease sensitivity to
MSPs. Thus, the ratio $\text{DM}/P$ is a measure of how
deep a survey probes the Galaxy for MSPs.  Modern pulsar surveys thus
compensate for these effects by narrow filterbank channels and short
sampling times \citep{2006ApJ...637..446C, 2010MNRAS.409..619K}.
Given the 3-MHz wide channels and slow sampling time of 250\,$\mu$s
of the PMPS data, the discovery of pulsars like PSR~J1652$-$48 and
PSR~J1748$-$3009 is surprising.

\begin{figure*}
  \begin{center}
    \ifcase \bwswitch
    \includegraphics[width=\columnwidth]{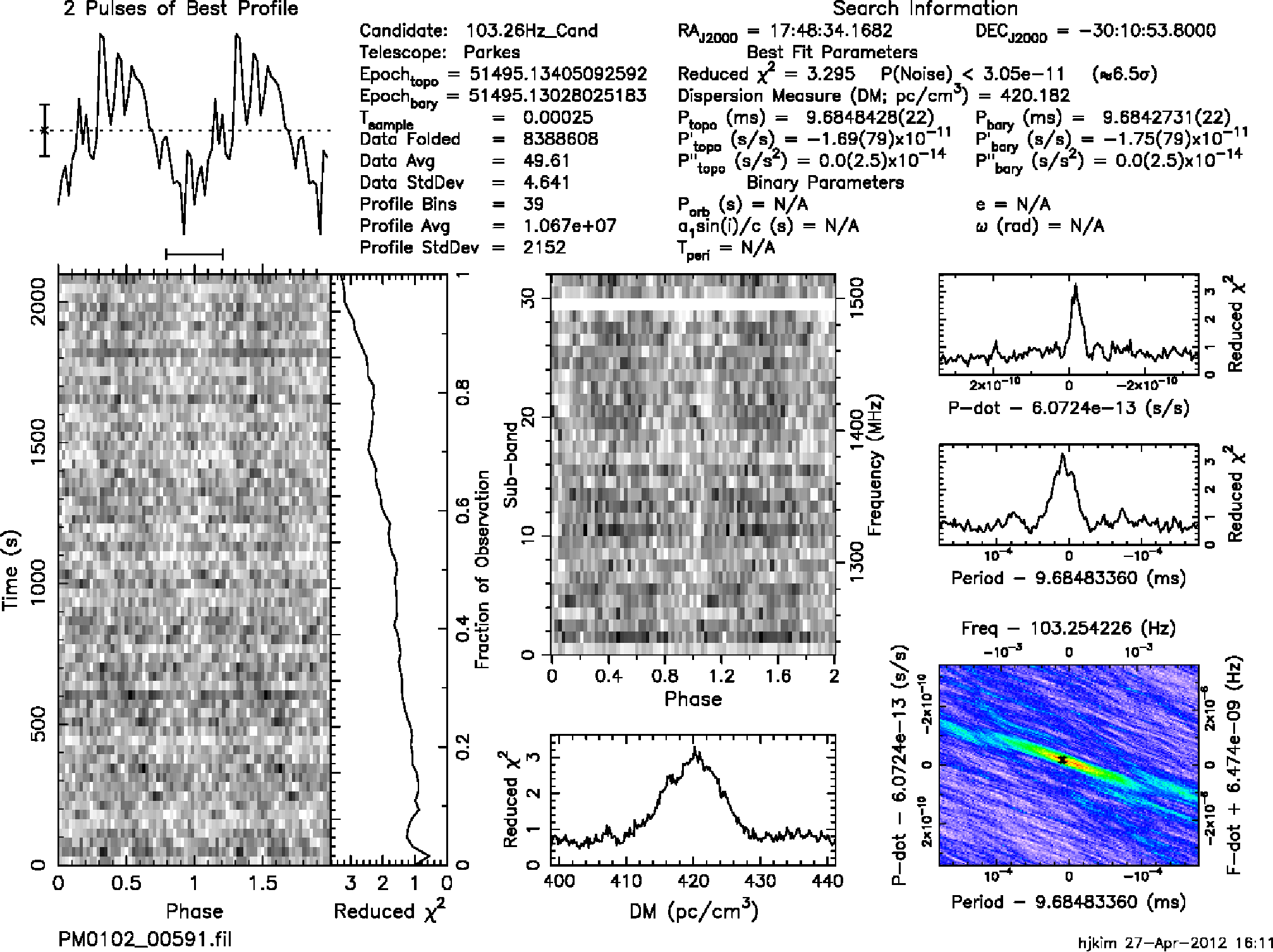}
    \hfill
    \includegraphics[width=\columnwidth]{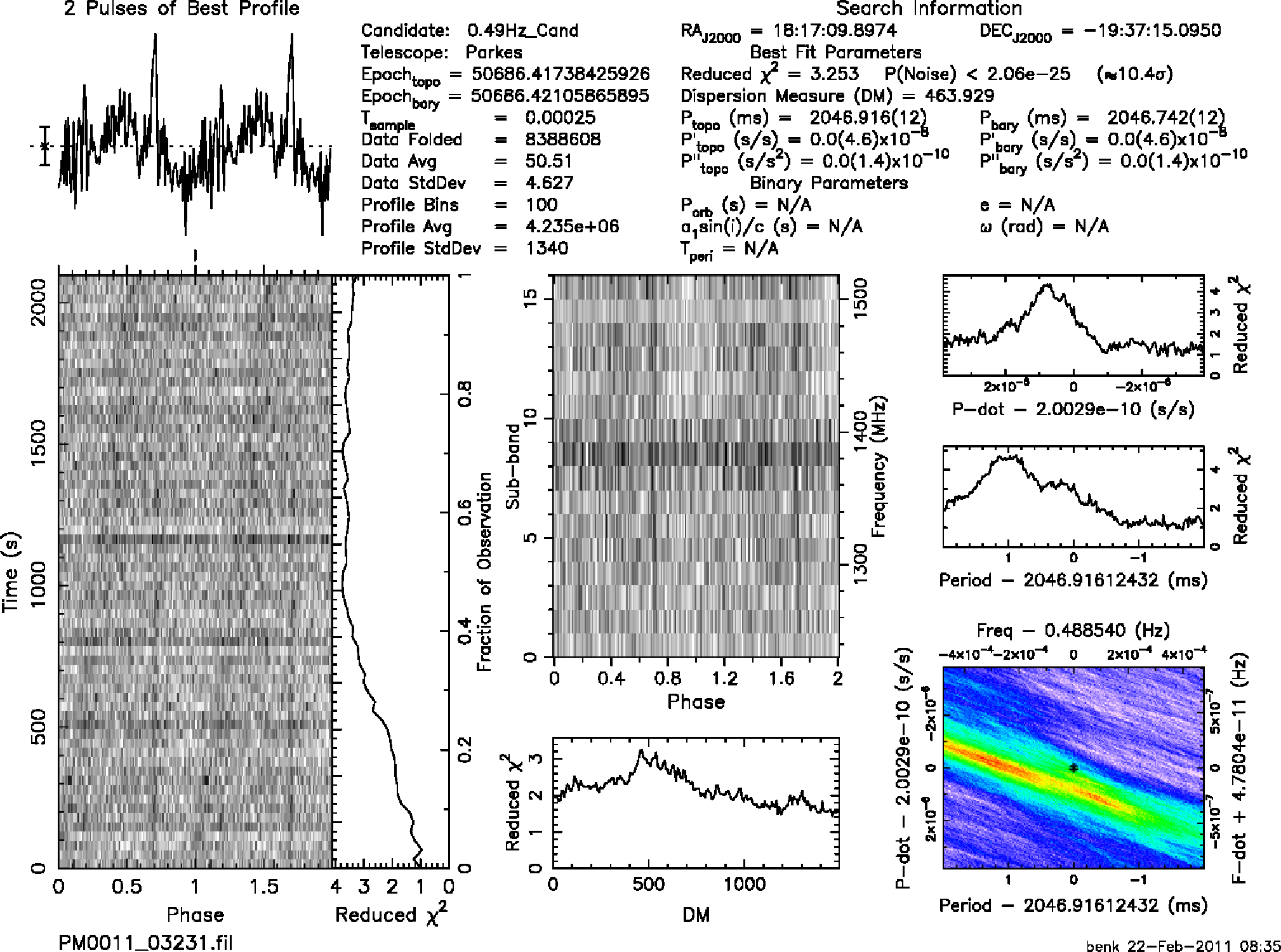}
    \or
    \includegraphics[width=\columnwidth]{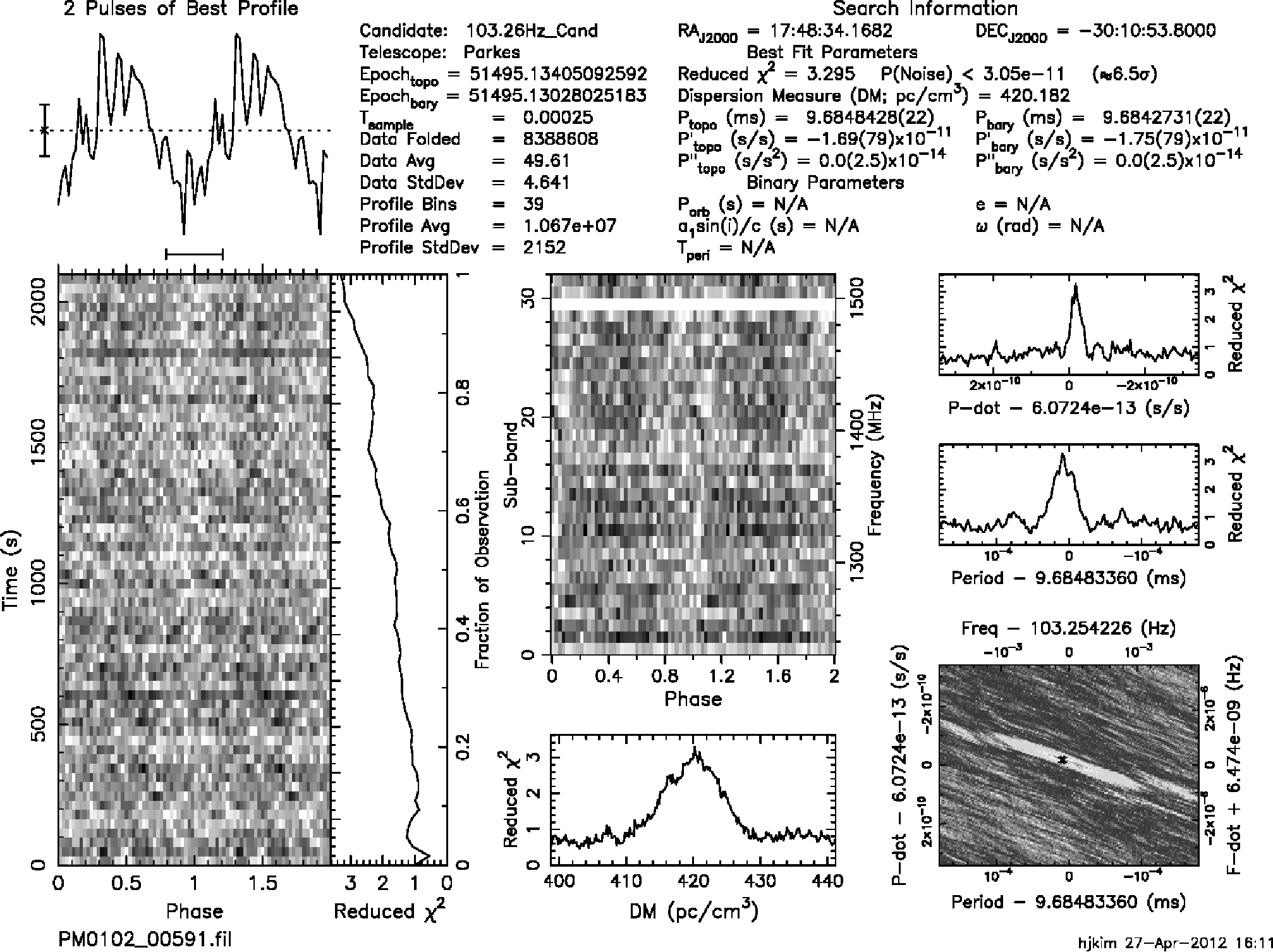}
    \hfill
    \includegraphics[width=\columnwidth]{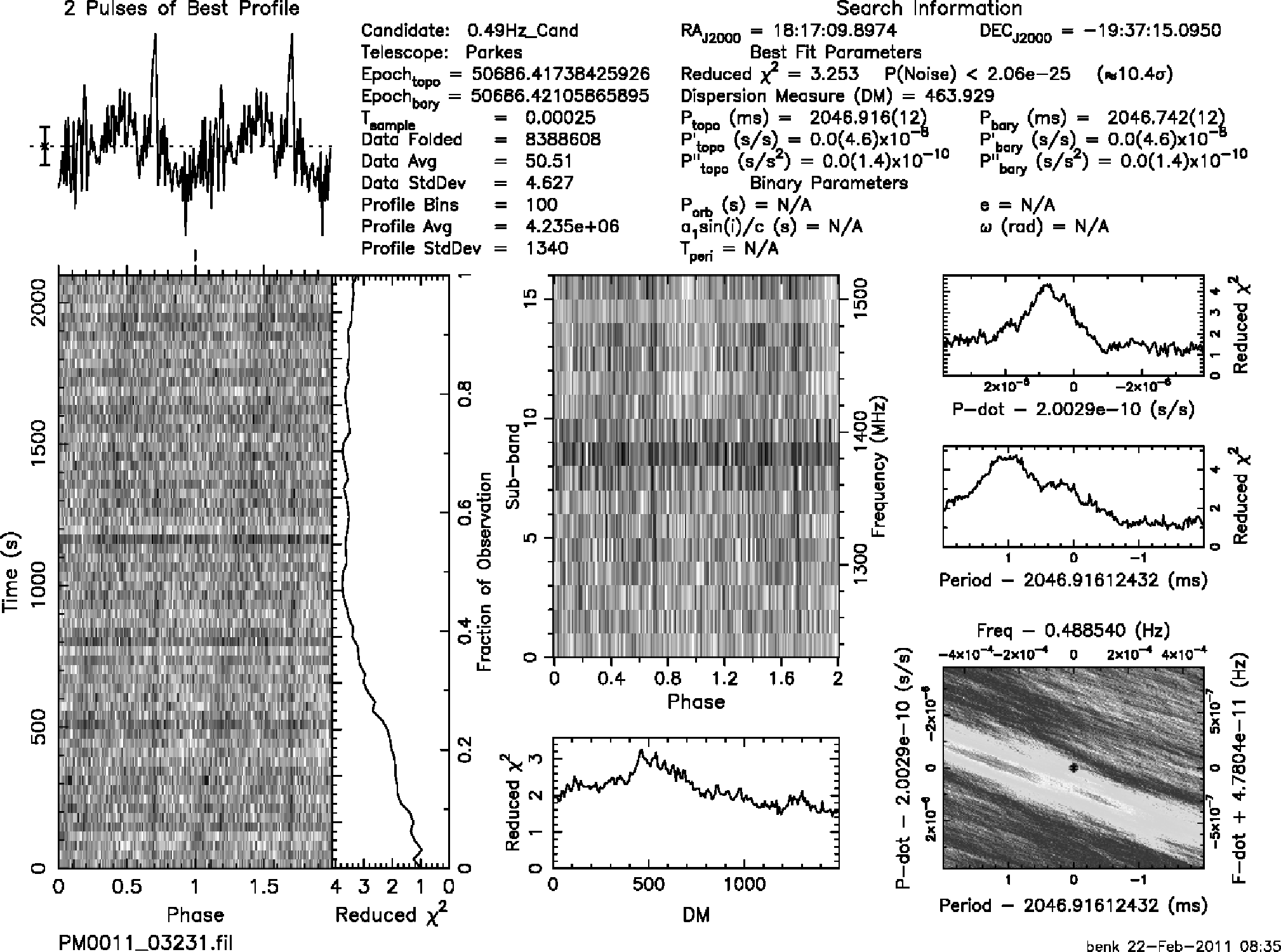}
    \fi
    \caption{\textsc{presto} discovery plots for two of the 24 pulsars found by
      \EAH{} in the PMPS data. \textbf{(Left)} The discovery plot of
      the millisecond pulsar J1748$-$3009.  Its pulse profile is very
      wide and appears almost sinusoidal from smearing over the 3\,MHz
      wide filterbank channels used in the PMPS. \textbf{(Right)} The
      discovery plot of the slow pulsar J1817$-$1938, which exhibits
      clear signs of intermittency. Its emission gradually disappears
      towards the middle of the 35\,min observation. Note also that
      the pulsar was discovered in our search despite the much
      stronger RFI at nearby periods, which is clearly visible. This
      discovery plot shows a slight phase-drift because it was folded
      with a small frequency offset.}
    \label{fig:discoveries}
  \end{center}
\end{figure*}

\begin{figure*}
  \begin{center}
    \includegraphics[width=\textwidth]{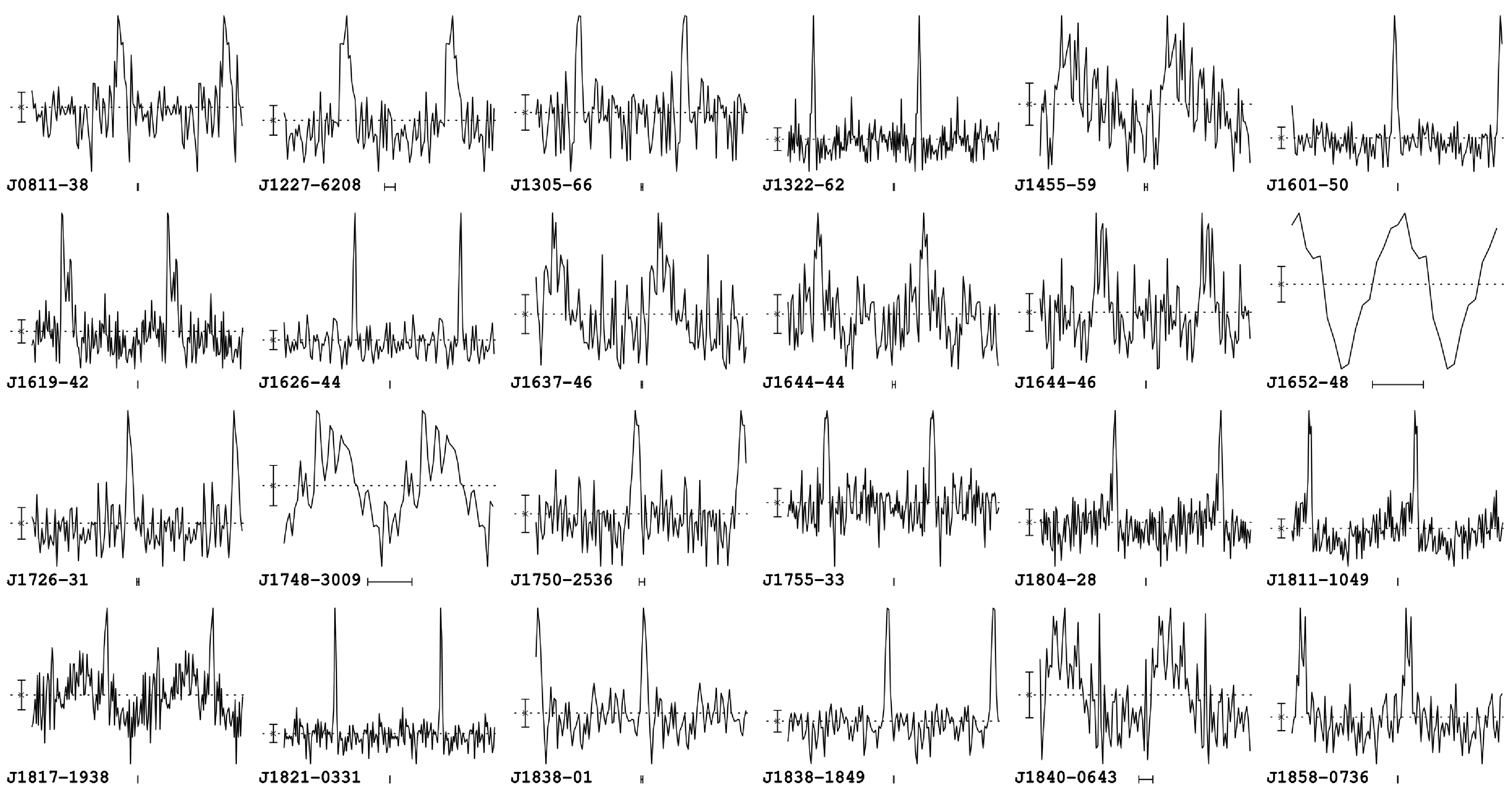}
    \caption{Discovery pulse profiles for all 24 pulsars. Each panel
      shows the flux density over two rotations of the pulsar, is
      based on a single 35\,min observation from the PMPS,
      and has been folded with \textsc{prepfold}. On the left side
      of each profile is a vertical bar showing the standard deviation
      of the noise baseline, and on the bottom is a horizontal bar
      showing the time resolution in the profile. Note the almost
      sinusoidal pulse profiles of PSR~J1652$-$48 and
      PSR~J1748$-$3009, which are strongly affected by filterbank
      channel smearing. The pulse profile of PSR~J1455$-$59 shows a
      pronounced multi-path scattering tail.}
    \label{fig:pprofiles}
  \end{center}
\end{figure*}

Tab.~\ref{tab:binary} lists the orbital parameters of three newly
discovered binary pulsars. Note that we have a coherent timing
solution for only one of the sources. Improved orbital parameters and
full timing solutions for the binary pulsars will be published in a
followup article. Fig.~\ref{fig:discoveries} shows the discovery
plots made with the \textsc{presto} tool \textsc{prepfold} for
two selected pulsars, while Fig.~\ref{fig:pprofiles} displays the
discovery pulse profiles of all 24 sources. We describe notable
sources in more detail below.

\begin{table*}
  \caption{\rm Orbital parameters of three binary pulsar systems discovered by
    the \EAH{} analysis of the PMPS data.}
  \setlength{\extrarowheight}{2pt}
  \label{tab:binary}
\begin{tabular*}{\textwidth}{@{\extracolsep{\fill}}lllcllcr}
  \hline
  \multicolumn{1}{c}{PSR} & \multicolumn{1}{c}{data timespan} & \multicolumn{1}{c}{$a\sin(i)$} & \multicolumn{1}{c}{$e$} & \multicolumn{1}{c}{$T_0$} & \multicolumn{1}{c}{$P_\text{orb}$} & \multicolumn{1}{c}{$\omega$} & \multicolumn{1}{c}{Epoch}\\
  & \multicolumn{1}{c}{(MJD)} & \multicolumn{1}{c}{(lt-s)} & & \multicolumn{1}{c}{(MJD)} & \multicolumn{1}{c}{(d)} & \multicolumn{1}{c}{(degs)} & \multicolumn{1}{c}{(MJD)} \\
  \hline

  J1748$-$3009 & 56055 -- 56174 & 1.32008(1) & -- & 56069.162075(7) & 2.9338198(4) & -- & 51495.13\\

  J1750$-$2536 & 56036 -- 56176 & 20.06096(5) & 0.000392(4) & 56069.27(3) & 17.141650(4) & 54.7(7) & 50593.78\\

  J1840$-$0643 &  55699 -- 56161 & 113.2(2) & -- & 56044.4(1) & 937.1(7) & -- & 55930.01\\

  \hline
\end{tabular*}
\end{table*}

\subsection*{PSR J1227$-$6208}
This pulsar in a binary system with a 6.7-day orbit was independently
discovered by \citet{2012ApJ...759..127M} and will be further
characterized by Thornton et al.\ (in prep.). Although its significance
$\mathcal S = 17.9$ in the \EAH{} pipeline is relatively high, it is
rather inconspicuous in the overview plots because of strong remnant
RFI. The automated post-processing successfully identified this
pulsar.

\subsection*{PSR J1322$-$62}
The discovery observation of this isolated 1045\,ms pulsar at
DM $\approx 734$\,pc\,cm$^{-3}$ exhibits signs of
intermittency. Follow-up observations confirmed this trend: we
observed the pulsar three times at MJD~55592.7 ($T \approx 2100$\,s),
MJD~55615.7 ($T \approx 590$\,s), and MJD~55648.6 ($T \approx
2100$\,s), respectively, with the Parkes telescope. We only detected
the pulsar in the second (shortest) observation. The non-detections
in the two other observations correspond to upper limits on the flux density
of $S \lesssim 0.07$\,mJy, as computed from, e.g., Appendix~2.6
in \citet{pulsarhandbook}. Here, and for all following upper limits
we assumed a signal-to-noise threshold of $8\sigma$, the observed pulsar
duty cycle, sky temperature at the pulsar sky position and system parameters
as given in, e.g., \citet{2001MNRAS.328...17M}.

The pulsar apparently shows long-term nulling or intermittent emission behavior
\citep{2006Sci...312..549K, 2012ApJ...758..141L}. Like the other intermittent pulsars
presented here, their large inferred distances mean that interstellar
scintillation is unlikely to be the cause of the observed intensity
variations; at this observing frequency scintles of width of only a
few MHz are expected and therefore average out over the 288\,MHz band.
Further observations are required to quantify this effect for
PSR~J1322$-$63.

\subsection*{PSR J1455$-$59}
This isolated 176\,ms pulsar showed a strongly pronounced scattering
tail in its pulse profile in the discovery observation. Its DM
$\approx 498$\,pc\,cm$^{-3}$. Further measurements of the
pulse shape at multiple radio frequencies could be used to study the
interstellar medium, by determining exact values of the
pulse-broadening time scales $\tau_d$. Assuming a delta-function pulse
shape, we determine $\tau_d=0.08(4)$\,s from the discovery
observation. This is in very good agreement with the measurements in
\citet{2004ApJ...605..759B}. More exact measurements of $\tau_d$ in
turn can be used to improve Galactic electron density models
\citep{2002astro.ph..7156C}, and to update existing empirical
relations between $\tau_d$ and the DM \citep{2004ApJ...605..759B}.

\subsection*{PSR J1652$-$48}
This binary MSP has a spin period of 3.78\,ms and a
DM of $\approx 188$\,pc\,cm$^{-3}$. This pulsar has the
fifth highest value of DM$/P = 49.7$\,pc\,cm$^{-3}$\,ms$^{-1}$ of all
Galactic field pulsars. As discussed above, the ratio is a measure of
the depth to which pulsar surveys probe our Galaxy for MSPs. The
value for PSR~J1652$-$48 is only surpassed by that for
PSR~J1903$+$0327, PSR~J1900$+$0308, and PSR~J1944$+$2236 found in the
PALFA survey \citep{2008Sci...320.1309C, 2012ApJ...757...90C}, and by
that for PSR~J1747$-$4036, found through a radio search of an
unidentified \textit{Fermi} source \citep{2012ApJ...748L...2K}.
The pulsar is being timed at the Parkes Observatory.

\subsection*{PSR J1726$-$31}
We confirmed PSR~J1726$-$31 upon the first re-detection attempt at
Parkes telescope in 2012 February. In nine subsequent observations at
Parkes, only one further detection of this pulsar at MJD~56111.5 has been
made. The other non-detections were done with the following integrations
times (upper limit on the flux density in parentheses): once with
$T\approx 3600$\,s ($S \lesssim 0.08$\,mJy), four times with
$T\approx 2100$\,s, once with each $T\approx 1800$\,s, $T\approx 1500$\,s,
and $T\approx 1300$\,s ($S \lesssim 0.1$\,mJy in each case). The large
distance of 4.1\,kpc, inferred from $\text{DM} \approx 264$\,pc\,cm$^{-3}$
means interstellar scintillation is unlikely to be the cause of the observed
intensity variations.

A search of the Parkes archival observation logs has revealed that
this pulsar was independently identified as a promising candidate in
1999, at the time of the original PMPS observations. Four confirmation
attempts were made in the same year, but without success. Three of them
have integration times $T\approx2100$\,s and each limits the flux density to
$S\lesssim 0.1$\,mJy, one has $T\approx360$\,s and therefore $S\lesssim 0.25$\,mJy.

Including our observations with these earlier confirmation attempts, and the
original survey observation, the pulsar has been observed for a total
of $\approx 8.6$\,hr. Of this time the pulsar is visible for
$\approx 1.9$\,hr, suggesting the pulsar is detectable about 20\,\% of the time.

In addition to the intermittency, there is evidence for large period
changes between observations, and one instance of significant
line-of-sight acceleration ($a \approx 11$\,m\,s$^{-2}$). This suggests
that this pulsar is a member of a compact binary system. Additional timing
observations to characterize this pulsar, its possible companion, and
the orbital parameters are ongoing.

If this pulsar is in a compact binary system, it is strikingly similar
to PSR~J1744$-$3922 \citep{2004MNRAS.355..147F}. PSR~J1744$-$3922 is
mildly recycled with a spin period of 172\,ms (cf.\ 123\,ms for
PSR~J1726$-$31) and resides in a compact binary system
($P_\text{orb}\approx 4.6$\,hr). Like our discovery, PSR~J1744$-$3922
also exhibits nulling and is visible for approximately 25\,\% of
the time at 1.4\,GHz.

PSR~J1744$-$3922, and thus PSR~J1726$-$31, might be the members of a new
class of binary pulsars as proposed by \citet{2007ApJ...661.1073B}:
these binary pulsars have long spin periods, large magnetic fields
($\sim10^{10-11}$\,G), low-mass companions, and low orbital
eccentricities. Their evolutionary history is not well understood and
known formation channels fail to explain all of their properties.

A different possible explanation for the observed intensity variations
are eclipses in a compact binary systems: ``black widow'' systems like
PSR~B1957$+$20 \citep{1988Natur.333..237F}. This seems unlikely,
though, since the spin period of PSR~J1726$-$31 is significantly
larger than those of other known eclipsing binary pulsars
\citep{2005ASPC..328..405F}.

\subsection*{PSR J1748$-$3009}
This pulsar is a 9.7\,ms pulsar in a 2.93\,day binary orbit. With
$\text{DM} \approx 420$\,pc\,cm$^{-3}$ it has by far the highest DM of all
MSPs known to date. Its value of DM$/P = 43.3$\,pc\,cm$^{-3}$\,ms$^{-1}$ is the
seventh highest value published \citep{2005yCat.7245....0M,
2012ApJ...757...90C}. The pulsar is currently being observed at Jodrell
Bank on a regular basis to determine a full timing solution, which we will
publish in a second paper on our search. Tab.~\ref{tab:binary} shows first
measurements of its orbital parameters. The mass function
$f\approx2.87\times10^{-4}$\,$M_\odot$ of this system indicates a
minimum (median) companion mass of 0.09\,$M_\odot$ (0.10\,$M_\odot$)
for a pulsar mass of 1.4\,$M_\odot$.

\subsection*{PSR J1750$-$2536}
This pulsar has a spin period of 34.7\,ms and was discovered
independently in two adjacent PMPS beams. It has a DM of
$\approx 178$\,pc\,cm$^{-3}$ and is a member of a binary system with an
orbital period of 17.1\,days. It is currently being observed regularly at
Jodrell Bank to improve the timing solution, which will be published
in a follow-up paper. Tab.~\ref{tab:binary} shows first measurements
of its orbital parameters. The mass function computed from the orbital
parameters is $f\approx 0.029$\,$M_\odot$, yielding a minimum (median)
companion mass of 0.47\,$M_\odot$ (0.56\,$M_\odot$) for a pulsar of
1.4\,$M_\odot$.

Besides the `common' low-mass binary pulsars (LMBPs) exists the rather
rare class of intermediate-mass binary pulsars (IMBPs). IMBPs differ
from the LMBPs by longer spin periods, more massive companions, and
larger orbital eccentricities. Further, their binary evolution
channels seem to be significantly different
\citep{2001ApJ...548L.187C}.
 
The combination of a long orbital period ($17.1$\,days), relatively long
spin period ($34.7$\,ms), and an eccentricity of
$e=3.92(4)\times10^{-4}$ makes PSR~J1750$-$2536 most likely an
IMBP. Plotting the orbital parameters of PSR~J1750$-$2536 in the
$P_\text{orb}$-$e$ plane clearly places this pulsar outside the
parameter space expected for LMBPs predicted by a
relation found by \citet{1992RSPTA.341...39P}, cf.\ Fig.~4 of
\citet{2001ApJ...548L.187C}. Its low inferred distance from the
Galactic plane $|z|=0.04$\,kpc is comparable to that of the other
known IMBPs.

We will determine the spin-down of the pulsar with an ongoing timing
campaign, from which we will in turn obtain the surface magnetic field of
the pulsar. If PSR~J1725$-$2536 is indeed an IMBP, its magnetic field
$B$ is expected to be relatively low at $B=(5-90)\times 10^8$\,G.

\subsection*{PSR J1817$-$1938}
This source was discovered by the \EAH{} project in 2011 February in
two independent PMPS beams and confirmed by a second observation in
2011 July. It has been independently discovered and published without
a timing solution by \citet{2012MNRAS.427.1052B}. We observed the
pulsar regularly at Jodrell Bank and obtained a timing solution, which
we report in Tab.~\ref{tab:all}. The discovery observation of this
2047-ms pulsar at DM $\approx 520$\,pc\,cm$^{-3}$ exhibited signs of
intermittency: it showed fading and re-appearing radio emission over
the course of both 35\,min discovery observations.  We re-observed
the pulsar with the Parkes telescope at MJD~55649.7 for $T\approx 1800$\,s
and performed a gridding observation the Effelsberg 100-meter telescope at
MJD~55725.0. The pulsar was not detectable in the Parkes observation
(thus $S \lesssim 0.09$\,mJy), but we found it permanently emitting
in an 11.5\,min observation with the Effelsberg telescope.

\subsection*{PSR J1821$-$0331}
This 902\,ms pulsar with DM $\approx 172$\,pc\,cm$^{-3}$ exhibited
signs of intermittency in its discovery observations. A follow-up
observation on MJD~55790.6 using the Parkes telescope did not
convincingly confirm the pulsar ($T\approx1800$\,s, $S \lesssim 0.06$\,mJy),
but a second observation on MJD~55808.2 at Jodrell Bank did. The
first confirmation attempt and varying flux density in the Jodrell
Bank observation confirm that this pulsar shows intermittent emission on
long time scales. Tab.~\ref{tab:all} shows its properties from a
fully-determined timing solution, obtained using observations with the
Lovell telescope at Jodrell Bank.

\subsection*{PSR J1840$-$0643}
This 35.6\,ms pulsar was identified independently in two PMPS survey
beams at DM $\approx 500$\,pc\,cm$^{-3}$. The barycentric spin period
observed in each one of these observations shows an increase that is
inconsistent with pulsar spin-down. Ongoing timing observations at
Jodrell Bank have revealed that this pulsar is in a binary system with
an unusually large orbital period of 937\,days (the fourth largest
known) and with a small eccentricity, $e\approx0$.
Tab.~\ref{tab:binary} shows measurements of its orbital parameters
from our coherent timing solution. The mass function $f\approx 1.77\times10^{-3}$\,$M_\odot$
for this system yields a minimum (median) companion mass of
0.16\,$M_\odot$ (0.19\,$M_\odot$) for a pulsar mass of 1.4\,$M_\odot$.

If the companion is a typical low-mass He white dwarf, as the mass
function suggests, plotting this pulsar on the Corbet diagram of spin
period versus orbital period \citep{1984A&A...141...91C} reveals that
this pulsar is located in the same region as other long orbital period
systems ($P_\text{orb} > 200$\,days) which contain recycled pulsars
with longer spin periods \citep{2011ASPC..447..285T}.
\cite{1999A&A...350..928T} provide an explanation for the origin of
such systems: the progenitor was likely a wide low mass X-ray binary
system. In such systems, there is only a short period of mass transfer
because the donor star is highly evolved by the time it fills its
Roche lobe.

If the companion is indeed a white dwarf this system is a potential
target for tests of the strong equivalence principle via the
Damour-Schaefer test \citep{1991PhRvL..66.2549D}. Unfortunately, it can be
shown that at least one of the fundamental criteria for this test is
not fulfilled: the angular velocity of the relativistic advance of
periastron, $\dot{\omega}$, must be appreciably larger than the
angular velocity of the systems rotation around the Galactic center,
$\Omega_\text{Gal}$. For this system, and
based on the DM-inferred distance, $\Omega_\text{Gal} \approx
3.6$\,deg\,Myr$^{-1}$, and $\dot{\omega} \approx 3.0$\,deg\,Myr$^{-1}$ (N.\ Wex, private
communication, 2012). If the system orientation is well understood, tests of
SEP can be made via other methods, however this also requires high
timing precision \citep{2012CQGra..29r4007F}.

\section{Discussion and Conclusions}\label{sec:discussion}
Our discoveries, and those of other ongoing analyses
\citep{2012ApJ...759..127M} demonstrate that the instrumental
sensitivity is not the limiting factor in searches of the PMPS data:
using new methods and more computing power for re-analyses of the
same data still yields new pulsar discoveries. Pushing into yet
unexplored regions of parameter space in future re-analyses might
lead to further discoveries, since the orbital parameter space in
our search was also limited by the available computing power, see
Sec.~\ref{subsec:searchspace}. Extending the search sensitivity to
the point where it is limited by the instrumental sensitivity is
critical for efforts to characterize the properties of the pulsar
population from surveys. Usually completeness is implicitly assumed
\citep{2006ApJ...643..332F} or characterized by simple metrics
\citep{2012arXiv1210.2746L}, when modeling the population. The PMPS,
despite being by far the most-analyzed pulsar survey, is not
`completed' -- there are numerous as-yet-unconfirmed candidates, both
binary and isolated \citep{ralphpaper}.

Although approximately one-third of all of the pulsars detected in the
PMPS are strongly detected in single pulse searches
\citep{2010MNRAS.401.1057K}, none of the 24~sources reported here show
single pulses stronger than $5\sigma$ (a typical noise-floor level for
a PMPS single pulse search) in the survey observations. This implies
that the `intermittency ratio', the ratio of peak single pulse to
folded signal-to-noise ratios, for each sources is at most
$\gtrsim 0.3$, meaning that these sources were at least two to six times
more detectable in a periodicity search than in a single pulse search
\citep{2010PhDT.......460K}. Despite this fact, three of the sources
were subsequently found to be intermittent (see
Section~\ref{sec:discoveries}) highlighting that pulsars can be
variable on a wide range of timescales \citep{2012arXiv1210.5397K}.

Additionally, intermittency has an impact on completeness, e.g., if a
pulsar is `on' for only 10\% of the time and its nulling periods are
longer than the survey integration length, there is only a 1\%
chance of detecting it in both the survey and in a confirmation
observation. From each of the past PMPS analyses, there are many
high-quality candidates which have never been confirmed and may belong
to such a category. The discovery of PSR~J1808$-$1517
\citep{ralphpaper}, which took many hours of re-observations to
confirm, emphasizes this point. To address this, future pulsar surveys
will scan the sky multiple times.

The spacing of the orbital templates in our parameter space ($a\sin
i$, $P_\text{orb}$, and orbital phase $\psi$) means that we have full
sensitivity (up to the nominal mismatch and with the exception of
candidates lost during post-processing due to human error) to pulsars
with spin frequencies $\leq 130$\,Hz, with increasingly reduced
sensitivity at higher spin frequencies. This explains in part why we
did not detect all of the MSPs recently reported by
\citet{2012ApJ...759..127M}, although we did independently discover
PSR~J1227$-$6208. Some of the MSPs from \citet{2012ApJ...759..127M}
were inside our frequency search range, but were not significant
enough to be seen in the overview plots and to `survive' the automated
post-processing stage. Others were detected only at their
sub-harmonics, because their spin frequencies are outside our search
range, with massively reduced significances, and suffered the same
fate. In one case, strong RFI masked the pulsar completely in our
analysis.

Another limitation on the completeness of our search is our assumption
of circular orbits. This means that we have reduced sensitivity to
eccentric systems but retain 50\% of the full sensitivity to binaries
with eccentricities of $e\approx 0.1$ (for a spin frequency of
130\,Hz) averaging over all possible orbital phases of detection
\citep{benthesis}. We note that our search is sensitive to signals at
lower spin frequencies with orbital parameters outside the parameter
space covered by the template bank.

Future searches on the PMPS and other pulsar survey data will be able
to expand the search to increasingly larger parts of the binary
parameter space at higher sensitivities, likely through volunteer
distributed computing. Three main effects can play important roles in
future searches and their design. First, Moore's Law is expected to
continue for at least another few years, further increasing the
computing power available through CPUs. Second, improvements in GPUs
add a new, powerful and ubiquitous computing resource to volunteer
distributed computing. Lastly, the characterization of the RFI
environment from our PMPS analysis could improve RFI mitigation in
future searches and increase the overall sensitivity.

How much could future searches profit from the increase in computing
power from Moore's Law and faster GPUs? To answer this, we
(conservatively) assume that the computing power grows by 30\% each
year. Over the coming decade, this would result in an increase of a
factor of 14 in computing power. As discussed in
Sec.~\ref{subsec:searchspace}, the number of orbital templates grows
with $f_\text{max}^3$, and the computing costs of the search code
(FFTs and harmonic summing) grow roughly with $f_\text{max}$.  A
factor 14 increase in computing power would thus enable \EAH{} to
search the PMPS data for spin frequencies up to $f_\text{max}\approx
250$\,Hz, increasing sensitivity to faster pulsars in compact binary
systems.  Alternatively, a follow-up analysis could extend the orbital
search parameter space to larger projected radii or to shorter orbital
periods.

We think it is remarkable that even with the large computing power
provided by \EAH{}, our search is still computationally limited. More
than a decade after the completion of the PMPS, the data still cannot
be analyzed with the highest possible sensitivity to relativistic
pulsars. One should be aware that the return of future analyses--measured
in the number of new discoveries--is likely going to become
smaller and smaller as an increasing fraction of the possible parameter
space is analyzed. Yet, the as of now missing parts of the parameter space
would contain the most relativistic pulsars, with high scientific potential
as described in Sec.~\ref{sec:introduction}.

Our results illustrate the capability of volunteer
distributed computing for the analysis of large astronomical data
sets, which will become increasingly important in the
future. Distributed computing projects could play an important role in
meeting the ever-growing need for computing power in data-driven
research projects.

In this paper, we have described in detail the \EAH{} pulsar search
algorithm, the post-processing analyses, the discovery of 24 new
pulsars, and the presentation of timing solutions for five of these
sources. In a future paper we will expound upon the implications for
the population estimates for merging binaries in the Galaxy, as this
is far beyond the scope of this article.

\section*{Acknowledgements}\label{sec:acknowledgements}
We thank all \EAH{} volunteers, especially those whose computers found
the pulsars with the highest statistical significance\footnote{Where
  the real name is unknown or must remain confidential we give the
  \EAH{} user name and display it in single quotes.}.  PSR~J0811$-$38:
the Atlas computer cluster at the Albert Einstein Institute, Hannover,
Germany and the Nemo computer cluster at the Department of Physics,
University of Wisconsin--Milwaukee, Milwaukee, USA. PSR~J1227$-$6208:
Rolf Schuster, Neu-Isenburg, Germany and Darren Chase, Adelaide,
Australia. PSR~J1305$-$66: Du{\v s}an Pirc, Dom{\v z}ale, Slovenia and
`Victor1st'. PSR~J1322$-$62: Vadim Gusev, Petrozavodsk, Russia and
David Mason, Leawood, USA. PSR~J1455$-$59: David Peters, Kiel, Germany
and James Drews (UW-Madison), Madison, USA. PSR~J1601$-$50: Sirko
Rosenberg, Bautzen, Germany and Ton van Born, Nieuw-Vennep,
Netherlands.  PSR~J1619$-$42: `Metod, S56RKO' and Peter Grosserhode,
Las Vegas, USA. PSR~J1626$-$44: Aku Leijala, Veikkola, Finland, and
`Og'. PSR~J1637$-$46: Riaan Strydom, South Africa and
`Edelgas'. PSR~J1644$-$44: Jesse Charles Wagner II [USA] and
`Ras'. PSR~J1644$-$46: Augusto Cortemiglia, Tortona, Italy, and
`Axiel'. PSR~J1652$-$48: Brian Adrian, Dade City, USA and `Craig
G'. PSR~J1726$-$31: Bogus\l{}aw Sobczak, Krakow, Poland and Steve
Mellor, Perth, Australia. PSR~J1748$-$3009: J\"urgen Sauermann,
Berlin, Germany and `Stan Galka'. PSR~J1750$-$2536:
Frederick~J.~Pfitzer, Phoenix, USA, and Benjamin Rosenthal Library,
Queens College, CUNY, Flushing, USA. Independent detection of
PSR~J1750$-$2536 in a second PMPS beam: `Masor\_DC' and
Gordon~E.~Hartmann, Dover, USA. PSR~J1755$-$33: `Omega Sector - Game
Systems' and Dwaine Maggart, Van Nuys, USA. Independent detection of
PSR~J1755$-$33 in a second PMPS beam: `revoluzzer' and `Jacek
Richter'.  PSR~J1804$-$28: Drew Davis, Urbandale, Iowa, and John-Luke
Peck, TerraPower \& Intellectual Ventures, Seattle,
USA. PSR~J1811$-$1049: Ingo Eberhardt, Gross-Zimmern, Germany and
`Paul Serban'.  PSR~J1817$-$1938: `Jaska' and Keith Sloan, Nr
Winchester, UK. Independent detection of PSR~J1817$-$1938 in a second
PMPS beam: Chris Sturgess, New Canaan, USA and `Companion\_Cube'.
PSR~J1821$-$0331: `Robert Hoyt' and Kevin Battaile, Bolingbrook,
USA. PSR~J1838$-$01: Eric Nietering, Dearborn, USA and `Tim Taylor'.
PSR~J1838$-$1849: `gwyll' and `IG\_the\_cheetah'. PSR~J1840$-$0643:
Terry Dudley, San Francisco, USA and Nemo (see above). Independent
detection of PSR~J1840$-$0643 in a second PMPS beam: Trey Todnem,
Tucson, USA and Nemo (see above).  PSR~J1858$-$0736: Christoph Donat,
Ingolstadt, Germany and `gone'.

This work was supported by the Max Plank Gesellschaft and by NSF
grants 1104902, 1105572, 1148523, and 0555655. The authors
thank Paulo Freire and Norbert Wex for useful discussions. The
authors also thank Emily Petroff, Sarah
Burke-Spolaor, Dan Thornton, and David Champion for observations performed at Parkes. E.F.K.\
acknowledges the FSM for support. The Parkes radio telescope is part
of the Australia Telescope National Facility which is funded by the
Commonwealth of Australia for operation as a National Facility managed
by CSIRO.  The 100\,m Effelsberg telescope is operated by the
Max-Planck-Institut f\"ur Radioastronomie.

\bibliographystyle{yahapj}
\nocite{*}
\bibliography{pmps_eath}

\begin{appendix}
  \section{Alternative Coordinates used in the
    Post-processing}\label{subsec:coordtrans}
  For the generation of the overview plots and the automated
  post-processing algorithms described in Secs.~\ref{sec:postproc} and
  \ref{subsec:autopostproc}, new parameter space coordinates are
  used. The detection statistics $\mathcal P_n$ introduce correlations
  between the physical coordinates of the parameter space.  These
  correlations can partly be resolved by switching to different coordinates.
  \citet{2008PhRvD..78j2005P} has shown the value of this method for
  the search for continuous gravitational waves. We apply a similar
  method here in the post-processing step of our analysis.

  Let us rewrite the phase model \eqref{eq:fullphase} in a polynomial
  expansion: \beq \Phi\left(t;\mathbf{\Lambda}\right) =
  2\pi\sum_{j=1}^\infty \nu_j t^j + \Phi_0 \quad\text{with}\quad
  \nu_1=f\left(1 +
    \frac{a\sin(i)}{c}\Omega_\text{orb}\cos\left(\psi\right)\right)
  \quad\text{and}\quad
  \nu_2=-\frac{a\sin(i)\Omega_\text{orb}^2\sin\left(\psi\right)}{2c}f,\label{eq:phasemodelapprox1}\eeq
  where the orbital parameters are defined as before. In the
  definition of $\nu_1$, the factor multiplying $f$ is always close to
  unity, with typical $a\sin(i)\Omega_\text{orb}/c \lesssim
  10^{-3}$. Thus, $\nu_1$ is not simply changing the scale of $f$, but
  is rather applying a shift.

  Now insert this expansion into the expected value of detection
  statistic, given in Eq.  (5) of \citet{2013arXiv1303.0028A}. The
  power in the $n'$th Fourier mode is given by \beq
  \langle \mathcal P_n\left(\mathbf{\Lambda},\mathbf{\Lambda}'\right)
  \rangle \approx \left| \frac{\mathcal A_n}{2}\right|^2 \left|
    \frac{1}{T}\int_0^T \!\!\!\!\dm\!t \,\exp\left[2\pi i n\left(
        \sum_{j=1}^\infty \left(\nu_j - \nu'_j\right) t^j
      \right)\right]\right|^2
\label{eq:approxdetstat}
\eeq The detection statistic $\mathcal P_n$ is maximal if the argument
of the exponential is zero for all $t$. This is the case if and only
if $\nu_j - \nu'_j = 0$ for all $j \in \mathbb{N}$, since
$\left(1,t,t^2,\dots \right)$ is a basis of the vector space of real
polynomials. Thus, the argument of the exponential can only be zero if
all coefficients differences vanish. This defines a family of
``hyper-surfaces'' $\nu_j - \nu'_j = 0$ introduced in
\citet{2008PhRvD..78j2005P}.  These hyper-surfaces describe the
correlations between points in the parameter space.  Points with a low
mismatch lie on the intersection of all hyper-surfaces.

\begin{figure}
    \begin{center}
      \ifcase \bwswitch
      \includegraphics[width=0.5\columnwidth]{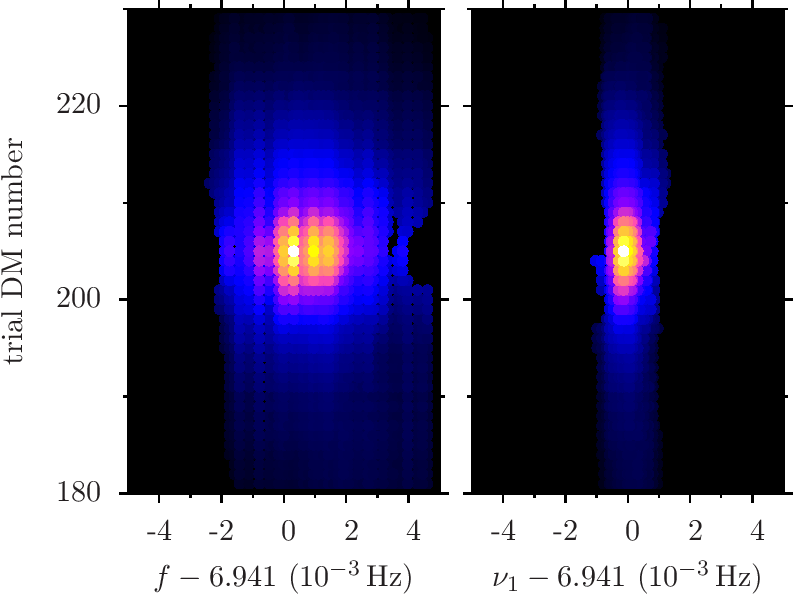}
      \or
      \includegraphics[width=0.5\columnwidth]{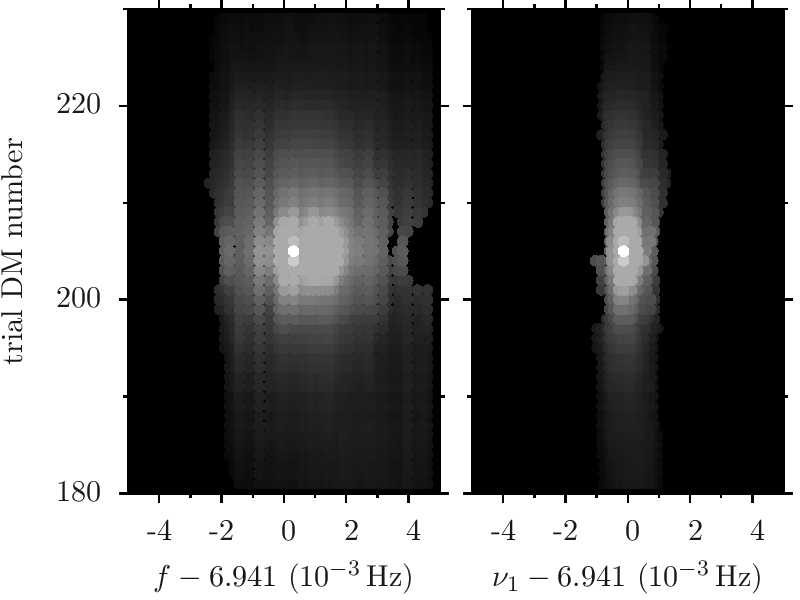}
      \fi
      \caption{The use of new plotting coordinates reduces the
        parameter spread, grouping physically related candidate
        signals closer together. \textbf{(Left)} Significance
        $\mathcal S$ as a function of spin frequency $f$ and
        dispersion measure DM for the detection of the relativistic
        pulsar J1906$+$0746 in the PMPS data. \textbf{(Right)}
        Significance $\mathcal S$ as a function of the spin frequency
        $\nu_1$ at the detector, and DM for the same data set.  Note
        the significant reduction of data point spread by about a
        factor four. This reduction is not due to a scaling but due to
        resolving correlations between the physical parameters, as
        discussed below Eq.~\eqref{eq:phasemodelapprox1}.}
      \label{fig:hypersurfacespread}
    \end{center}
  \end{figure}

Switching to the coordinates $\nu_1$ and $\nu_2$ moves templates
triggered by the same physical signal closer together, because these
coordinates resolve the correlations between the physical parameters
introduced by the detection statistic.
Fig.~\ref{fig:hypersurfacespread} demonstrates this effect by showing
both the spread of \EAH{} results for the detection of the
relativistic pulsar J1906$+$0746 in the PMPS data.

  Effectively, $\nu_1$ defines a the signal frequency at the detector
  at the beginning of the observation. The second coefficient,
  $\nu_2$, is proportional to the change of that frequency. This can
  be seen from taking the partial time derivative
  $1/(2\pi) \partial/\partial t$ of Eq.~\eqref{eq:phasemodelapprox1}
  \beq f(t) = \sum_{j=1}^\infty j\nu_j t^{j-1} = \nu_1 + 2\nu_2 t
  +\cdots \eeq
\end{appendix}

\end{document}